\documentclass[useAMS,usegraphicx,usenatbib]{mn2e}

\usepackage{graphicx}
\usepackage{psfig}

\title[Updating the cosmology of your halo catalogue]{How accurate is it to update the cosmology of your halo catalogues?}

\author[Andr\'es N. Ruiz et al.]{Andr\'es N. Ruiz$^{1,2,3}$\thanks{E-mail:
andresnicolas@oac.uncor.edu}, Nelson D. Padilla$^{2,3}$, Mariano J. Dom\'{\i}nguez$^{1,4,6}$, Sof\'ia A. Cora$^{5,6}$\\
$^{1}$Instituto de Astronom\'{\i}a Te\'orica y Experimental (CCT C\'ordoba, CONICET, UNC), Laprida 922, C\'ordoba, X5000BGT, Argentina\\
$^{2}$Departamento de Astronom\'{\i}a y Astrof\'{\i}sica, Pontificia Universidad Cat\'olica, Av. Vicu\~na Mackenna 4860, Santiago, Chile\\
$^{3}$Centro de Astro-Ingenier\'\i a, Pontificia Universidad Cat\'olica, Av. Vicu\~na Mackenna 4860, Santiago, Chile\\
$^{4}$Observatorio Astron\'omico de C\'ordoba, Universidad Nacional de C\'ordoba, Laprida 854, C\'ordoba, X5000GBR, Argentina\\
$^{5}$Instituto de Astrof\'isica de La Plata (CCT La Plata, CONICET, UNLP), Observatorio Astron\'omico, Paseo del Bosque, B1900FWA, \\
La Plata, Argentina.\\
$^{6}$Consejo Nacional de Investigaciones Cient\'ificas y T\'ecnicas (CONICET), Rivadavia 1917, Buenos Aires, Argentina.}

\begin{document}

\date{Accepted 2011 August 13. Received 2011 July 28; in original form 2010 October 14}

\pagerange{\pageref{firstpage}--\pageref{lastpage}} \pubyear{2011}

\maketitle

\label{firstpage}

%%%%%%%%%%%%%%%%%%%%%%%%%%%%%%%%%%%%%%%%%%%%%%%%%%%%%%%%%%%%%%%%%%%%%%%%%%%%%%%%%%%%%%%%%%%%%%%%%
%%%%%%%%%%%%%%%%%%%%%%%%%%%%%%%%%%%%%%%%%%%%%%%%%%%%%%%%%%%%%%%%%%%%%%%%%%%%%%%%%%%%%%%%%%%%%%%%%

\begin{abstract}
We test and present the application of the full rescaling method by Angulo \& White (2010) 
to change the cosmology of halo catalogues in numerical simulations for cosmological parameter search 
using semi-analytic galaxy properties. We show that a reduced form of the method can be applied
in small simulations with box side of $\sim 50$ h$^{-1}$Mpc or smaller without loss of accuracy. We 
perform statistical tests on the accuracy of the properties of rescaled individual haloes, and also on 
the rescaled population as a whole. We find that individual positions and velocities are recovered with 
almost no detectable biases, but with a scatter that increases slightly with the size of the simulation 
box when using the full method. The dispersion in the recovered halo mass does not seem to depend on the 
resolution of the simulation. Regardless of the halo mass, the individual accretion histories, spin parameter 
evolution and fraction of mass in substructures are remarkably well recovered. In particular, in order 
to obtain a more accurate estimate of the halo virial mass, it was necessary to apply an additional correction 
due to the change of the virial overdensity and the estimate of its effect on a NFW virial mass. The 
mass of rescaled haloes can be underestimated (overestimated) for negative (positive) variations of 
either $\sigma_8$ or $\Omega_m$, in a way that does not depend on the halo mass. Statistics of abundances 
and correlation functions of haloes show also small biases of $<10$ percent when moving away from the base 
simulation by up to $2$ times the uncertainty in the WMAP7 cosmological parameters. The merger tree 
properties related to the final galaxy population in haloes also show small biases; the time since the 
last major merger, the assembly time-scale, and a time-scale related to the stellar ages show correlated 
biases which indicate that the spectral shapes of galaxies would only be affected by global age changes 
of $\sim150$Myr, i.e. relatively small shifts in their broad-band colours. We show some of these biases  
for different separations in the cosmological parameters with respect to the desired cosmology so that 
these can be used to estimate the expected accuracy of the resulting halo population. We also present 
a way to construct grids of simulations to provide stable accuracy across the $\Omega_m$ vs. $\sigma_8$ 
parameter space.
\end{abstract}


%%%%%%%%%%%%%%%%%%%%%%%%%%%%%%%%%%%%%%%%%%%%%%%%%%%%%%%%%%%%%%%%%%%%%%%%%%%%%%%%%%%%%%%%%%%%%%%%%
%%%%%%%%%%%%%%%%%%%%%%%%%%%%%%%%%%%%%%%%%%%%%%%%%%%%%%%%%%%%%%%%%%%%%%%%%%%%%%%%%%%%%%%%%%%%%%%%%

\section{Introduction}

The Lambda Cold Dark Matter ($\Lambda$CDM) cosmological model is the standard theoretical framework
for structure formation in the Universe. In order to understand how galaxies form and evolve in this 
cosmological context, we must also understand the properties of dark matter haloes over a wide range 
of physical scales and across the cosmic history. Numerical simulations provide one of the best methods 
for approaching this problem.

The cosmological parameters that provide the best match between the $\Lambda$CDM cosmology and several 
observations, are most often obtained using measurements of the power spectrum of temperature fluctuations 
in the cosmic microwave background \citep{sanchez06, hinshaw09, dunkley09, jarosik10} or of density 
fluctuations in the galaxy distribution \citep{sanchez06, percival07}; but these are limited to the linear 
regime of density fluctuations. In order to make comparisons between the model and observations in 
the non-linear regime, without making simplified assumptions such as the spherical or ellipsoidal 
collapse \citep{sheth01}, one would need to use fully non-linear numerical simulations.  However, these 
are too expensive in terms of computational time. Great efforts go into running even single simulations 
corresponding to one set of cosmological parameters, which can still be compared with a very wide range of 
observational measurements.

For the past several years the Millennium Simulation \citep{mill} has been the focus  for many studies 
of the distribution and statistical properties of dark matter haloes and provides the basis for the 
implementation of semi-analytic models of the evolving galaxy population \citep{galform06, croton06}. However, 
it is important to appreciate that this simulation was run using the WMAP1 cosmological parameter set 
\citep{wmap1}, which are rather different from the current best fit parameters. More recently the Bolshoi 
Simulation \citep{bolshoi} used cosmological parameters consistent with the latest WMAP5 
\citep{hinshaw09, komatsu09, dunkley09} and WMAP7 \citep{jarosik10} values. The main difference is that the
Millennium simulation used a substantially larger amplitude of perturbations than the Bolshoi one.
 
The differences in the parameters of the cosmological model can affect the abundances and properties 
of the  dark matter haloes at a given redshift. Consequently, the properties of the galaxies hosted 
by such haloes may also be affected. As a consequence the interpretation of many observational statistical 
tests, like those measuring clustering and luminosity functions, becomes more difficult. Notice that 
according to the cosmological parameter constraints from WMAP5 and cluster abundances \citep{rozo09}, 
the Millennium Simulation is about $2-\sigma$ away from the best fit model (see Figure $1$ in \citealt{bolshoi}). 
The small variations of the parameters of the background cosmological model also become important in the 
understanding of the baryonic processes involved in galaxy formation. An extreme case can be seen in 
\cite{cole94} where the maximum of the stellar formation rate (SFR) activity is found at $z=1$ due to 
their choice of $\Omega=1$, which favours mergers at low redshifts. 

\begin{small}
\begin{table*}
\caption{Relevant parameters used in the simulations. The columns show the name of the simulation, 
the number of particles, the gravitational softening, the initial redshift, the particle mass, 
the boxsize, the matter density parameter, the baryon density parameter, the primordial spectral index, 
the dimensionless Hubble parameter, and the linear fluctuations amplitude in spheres of 8$h^{-1}$Mpc. The 
cosmological constant density parameter is $\Omega_\Lambda = 1 - \Omega_m$ (flat models) in all cases.
The boxsize of the $B$ simulations were calculated using the resulting scaling factor $s$ 
obtained from the rescaling technique.}
\begin{center}
\begin{tabular}{ccccccccccc}
\hline
\hline
 Name & $N_{p}$ & $\epsilon$    & $z_i$ & $M_{p}$             & $L_{box}$     & $\Omega_{m}$ & $\Omega_{b}$ & $n$ & $h$ & $\sigma_8$  \\
      &         & [$h^{-1}$kpc] &       & [$h^{-1}$M$_\odot$] & [$h^{-1}$Mpc] &              &              &     &     &             \\
\hline
$A$    & $256^3$ & $5$  & $74.7$ & $8.93\times10^{8}$  & $60.00$  & 0.25 & 0.0450 & 1.00 & 0.73 & 0.90  \\
$Alow$ & $128^3$ & $15$ & $59.7$ & $7.15\times10^{9}$  & $60.00$  & 0.25 & 0.0450 & 1.00 & 0.73 & 0.90 \\
$Abig$ & $256^3$ & $30$ & $43.3$ & $9.08\times10^{10}$ & $280.00$ & 0.25 & 0.0450 & 1.00 & 0.73 & 0.90 \\
$B$    & $256^3$ & $5$  & $61.2$ & $1.38\times10^{9}$  & $67.68$  & 0.27 & 0.0469 & 0.95 & 0.70 & 0.82 \\
$Blow$ & $128^3$ & $15$ & $49.1$ & $1.11\times10^{10}$ & $67.68$  & 0.27 & 0.0469 & 0.95 & 0.70 & 0.82 \\
$Bbig$ & $256^3$ & $30$ & $36.3$ & $1.27\times10^{11}$ & $305.33$ & 0.27 & 0.0469 & 0.95 & 0.70 & 0.82 \\
\hline
$Bo_{+2} ~(\Omega_m + 2\sigma)$           & $128^3$ & $20$ & $55.7$ & $6.39\times10^{9} $ & $52.72$ & 0.33 & 0.0469 & 0.95 & 0.70 & 0.82 \\
$Bo_{+1} ~(\Omega_m + 1\sigma)$           & $128^3$ & $20$ & $52.5$ & $8.27\times10^{9} $ & $59.29$ & 0.30 & 0.0469 & 0.95 & 0.70 & 0.82 \\
$Bo_{-1} ~(\Omega_m - 1\sigma)$           & $128^3$ & $20$ & $45.7$ & $1.55\times10^{10}$ & $78.78$ & 0.24 & 0.0469 & 0.95 & 0.70 & 0.82 \\
$Bo_{-2} ~(\Omega_m - 2\sigma)$           & $128^3$ & $20$ & $42.2$ & $2.34\times10^{10}$ & $94.48$ & 0.21 & 0.0469 & 0.95 & 0.70 & 0.82 \\
$Bs_{+2} ~(\sigma_8 + 2\sigma)$           & $128^3$ & $20$ & $52.8$ & $1.11\times10^{10}$ & $67.68$ & 0.27 & 0.0469 & 0.95 & 0.70 & 0.88 \\
$Bs_{+1} ~(\sigma_8 + 1\sigma)$           & $128^3$ & $20$ & $50.9$ & $1.11\times10^{10}$ & $67.68$ & 0.27 & 0.0469 & 0.95 & 0.70 & 0.85 \\
$Bs_{-1} ~(\sigma_8 - 1\sigma)$           & $128^3$ & $20$ & $47.3$ & $1.11\times10^{10}$ & $67.68$ & 0.27 & 0.0469 & 0.95 & 0.70 & 0.79 \\
$Bs_{-2} ~(\sigma_8 - 2\sigma)$           & $128^3$ & $20$ & $45.5$ & $1.11\times10^{10}$ & $67.68$ & 0.27 & 0.0469 & 0.95 & 0.70 & 0.76 \\
$Bo_{+1}s_{+1} ~(\Omega_m+1,\sigma_8 + 1\sigma)$ & $128^3$ & $20$ & $54.4$ & $8.27\times10^{9} $ & $59.29$ & 0.30 & 0.0469 & 0.95 & 0.70 & 0.85 \\

\hline
\hline
\end{tabular}
\end{center}
\label{tab:params}
\end{table*} 
\end{small}

Many of the differences between the Millennium and Bolshoi simulations can be re-scaled using 
different methods.  A first attempt to change the cosmology of a simulation was presented by 
\cite{zheng02} which was later applied by \cite{harker07} with the aim of constraining the 
cosmological parameter related to the amplitude of linear fluctuations in spheres of $8$$h^{-1}$Mpc, 
$\sigma_8$, using the {\small GALFORM} semi-analytic model \citep{galform05} and only two individual 
dark matter simulations.  This approach already greatly diminished the computational time that 
would have involved running numerical simulations for each set of cosmological parameters. More 
recently, the method suggested by \cite{aw10} allows a more flexible change in the cosmological 
parameters of a simulation without incurring in a important loss of precision. This algorithm scales 
the output of a cosmological $N$-body simulation carried out for one specific set of cosmological 
parameters so that it represents the growth of structure in a different cosmology. The accuracy
of the rescalings can only be estimated at first order using extended Press-Schechter theory or, 
more accurately, by running numerical simulations and testing for particular characteristics.

However, the algorithm developed by AW10 is applied to every particle in the simulation and the post-processing 
of the outputs of the simulation must be repeated (i.e. identification of the haloes and construction of 
the merger trees). This can be computationally demanding, particularly in the case when the rescaling 
needs to be done several times. We are interested in studying whether applying the method to dark-matter 
haloes instead of to particles produces similar accuracies in the rescaled simulations, since this would 
already considerably reduce the computational time.

We study whether the use of the AW10 method allows to explore variations in the phenomenology of non-linear 
density fluctuations in the cosmological parameter space via, for example, monte-carlo sampling. This is the 
reason why we will concentrate on reducing as much as possible the computational time in obtaining a catalogue 
of haloes from an $N$-body simulation on a different cosmology using AW10, and also on obtaining even a 
reduced version of their method. In any of these cases, the computational time involved is dramatically 
reduced in comparison to what would be needed to run a complete simulation for each new set of cosmological 
parameters. In the case of the reduced algorithm, it can be applied to small simulations where linear 
corrections span scales of the order of the box size. In the full and reduced cases, the approach we follow 
can be used when the particle information is not available as will be the case of future large simulations. 
We perform a number of tests using $N$-body simulations in similar cosmologies than those corresponding to 
the Millennium and Bolshoi simulations, in order to quantify the accuracy of the recovery of individual and 
statistical properties of the dark matter haloes and also exploring the effect of different extrapolation baselines.

The outline of this paper is the following. In Section \ref{sec:method} we present a brief description of the 
rescaling technique presented by AW10. We also describe in this section the $N$-body simulations used in this 
work, and the aplicability of a reduced version of the method. The results of the statistical and individual 
properties of rescaled dark matter haloes are shown in Section \ref{sec:results}. In Section \ref{sec:extrapolation} 
we measure how the resulting halo catalogues are affected by the distance in the cosmological parameter plane 
$\Omega_m - \sigma_8$ around the base WMAP7-$\Lambda$CDM model; this can later be used to evaluate how precise 
is any rescaled dark matter halo catalogue to ensure optimal explorations of the cosmological parameter space. 
Finally, the conclusions are presented in Section \ref{sec:conclusions}.

%%%%%%%%%%%%%%%%%%%%%%%%%%%%%%%%%%%%%%%%%%%%%%%%%%%%%%%%%%%%%%%%%%%%%%%%%%%%%%%%%%%%%%%%%%%%%%%%%
%%%%%%%%%%%%%%%%%%%%%%%%%%%%%%%%%%%%%%%%%%%%%%%%%%%%%%%%%%%%%%%%%%%%%%%%%%%%%%%%%%%%%%%%%%%%%%%%%

\section[scaltec]{Scaling the halo catalogues of $N$-body simulations}
\label{sec:method}

\begin{figure}
\centering
\includegraphics[scale=0.4]{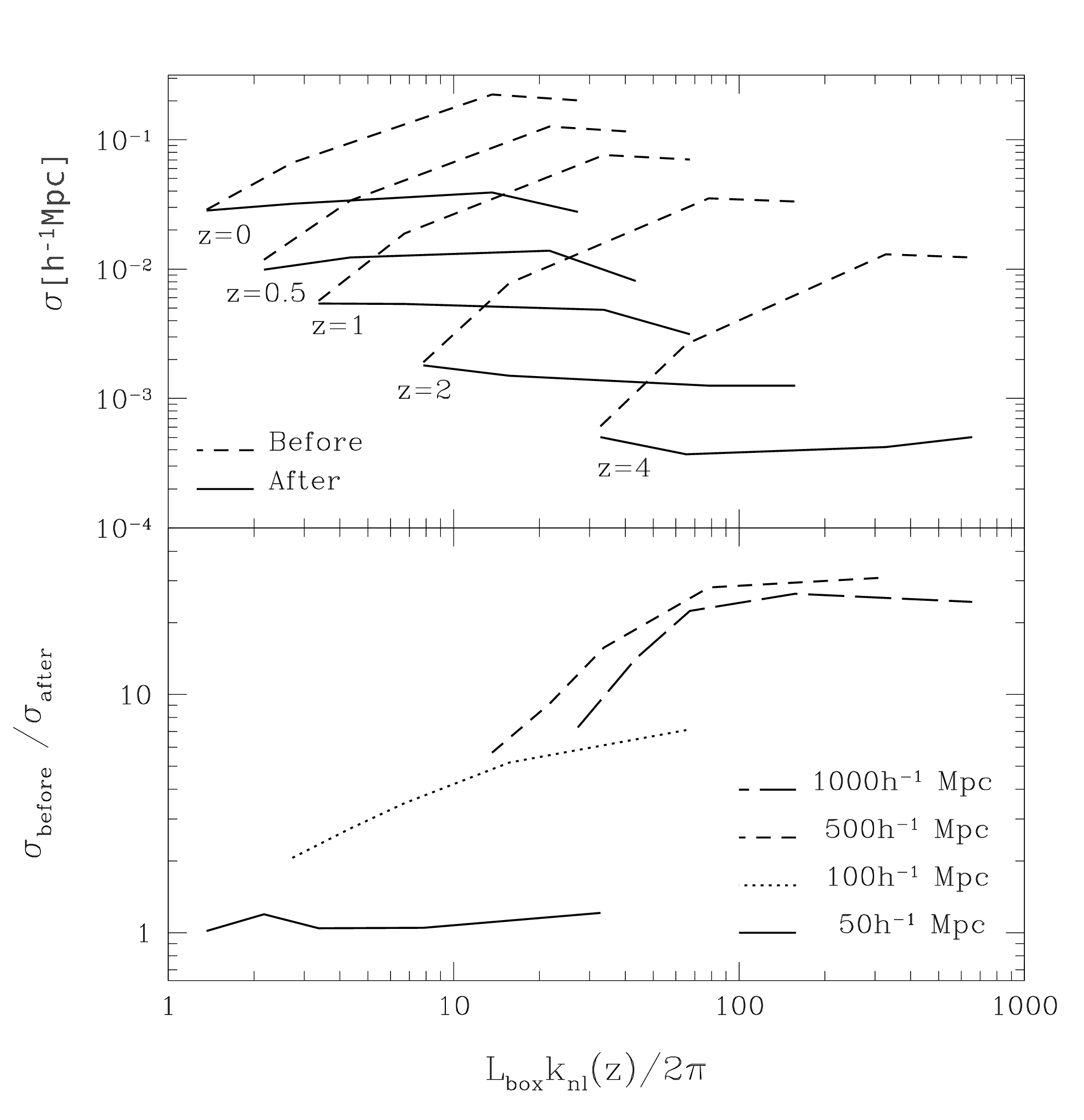}
\caption{Top panel: 1D rms difference in the rescaled particle positions 
for the full (solid) and reduced (dashed) AW10 method. The lines show the variation in the errors for 
fixed redshifts, as a function of the simulation box size; the pairs of solid and dashed lines correspond 
to different redshifts. Bottom panel: ratio between the 1D rms difference in the particle positions 
obtained after applying the reduced and full AW10 method; each line shows the results when fixing the 
box size but allowing the redshift to vary (box sizes are shown in the figure key). The x-axis 
shows the product of the box side of simulations $L_{box}$ and the non-linear limit mode $k_{nl}(z).$}. 
\label{fig:sigmas}
\end{figure}

We now briefly describe the procedure used to scale halo catalogues from a given cosmology into a new 
set of cosmological parameters. This procedure consists on the method presented by AW10 which can either 
be applied to haloes or individual particles in the simulation. If $P(k)$ is the linear matter power 
spectrum at $z=0$ we can define the variance of the linear density field as,

\begin{equation}
\sigma^2(R,z) = {D(z)^2 \over 4\pi} \int_{0}^{\infty} {k^2 P(k) W^2(kR)dk},
\label{sigma}
\end{equation}
where $R$ is a comoving smoothing scale, $D(z)$ is the linear growth factor normalised so that $D(z=0) = 1$ 
and $W(x)$ is the Fourier transform of a spherical top-hat filter defined by,

\begin{equation}
W(x) = 3{{\sin{(x)} - x \cos{(x)}} \over x^3}.
\end{equation}
Assuming that we want a halo catalogue of a given cosmology (denoted with $B$) evolved to a final 
redshift $z^f_B$ starting from another halo catalogue which has different cosmological parameters 
(denoted with $A$), the procedure is to find a length scaling $s$ of the boxsize, and a final redshift 
$z^f_A$ defined so that the linear fluctuation amplitude $\sigma_A (s^{-1}R,z_A^f)$ over the range 
$[s^{-1}R_1,s^{-1}R_2]$ is as close as posible to $\sigma_B (R,z_B^f)$ over the range $[R_1,R_2]$. 
This is done by minimising the function

\begin{equation}
\delta_{rms}^2 = {1 \over \ln(R_2/R_1)} \int_{R_1}^{R_2} {\left[1-{\sigma_A (s^{-1}R,z_A^f) 
               \over \sigma_B (R,z_B^f)} \right]^2 {dR \over R}},
\label{eq:drms}
\end{equation}
over $s$ and $z_A^f$. The values of $R_1$ and $R_2$ are selected so that $M(s^{-1}R_2)$ is the mass 
of the largest halo in the original simulation and $M(s^{-1}R_1)$ the mass of the smallest one.

Once we have the two parameters $s$ and $z_A^f$, the boxsize is scaled so that $L_B = sL_A$ and the 
earlier redshifts in the new cosmology ($z_B < z^f_B$) are obtained from those in the original cosmology 
($z_A < z^f_A$) through

\begin{equation}
D_B(z_B) = D_A(z_A) {D_B(z_B^f) \over D_A(z_A^f)}.
\label{outputs}
\end{equation}
After scaling the positions with $s$, and having matched the cosmological times, we must consider that 
the velocity and mass of the haloes need to be corrected using
\begin{equation}
\vec{v}_B = s {(1+z_A) \over (1+z_B)}{\dot{D}_B(z_B) \over \dot{D}_A(z_A)}{h_A \over h_B} \vec{v}_A,
\end{equation}
\begin{equation}
M_B = s^3 {\Omega_B \over \Omega_A} M_A,
\label{eq:mass_cor}
\end{equation}
where the dot indicates a derivate with respect to time, $h$ is the dimensionless Hubble parameter 
and $\Omega_B$ and $\Omega_A$ are the values of the matter density parameter $\Omega_m = \Omega_b + 
\Omega_{dm}$ for different cosmologies, being  $\Omega_b$ and $\Omega_{dm}$ the baryon and dark matter 
density parameters, respectively.

The steps presented in equations (1) to (6) will be referred to as the reduced AW10 method. The 
full implementation requires to correct the positions and velocities for residual differences
in the power spectrum of the two cosmologies and/or for the length scaling. To do this it is 
necessary to modify the contribution of the long wavelength components on the position and velocity 
fields using the Zel'dovich approximation. The range of modes were the correction is applied is 
$k < k_{nl}$, where $k_{nl}$ satisfies $\Delta^2(k_{nl})=1$ being $\Delta^2(k) = k^3P(k)/2\pi^2$ 
(See AW10 for the full details on this correction).

\begin{figure*}
\begin{picture}(400,250)
\put(-50,130){\psfig{file=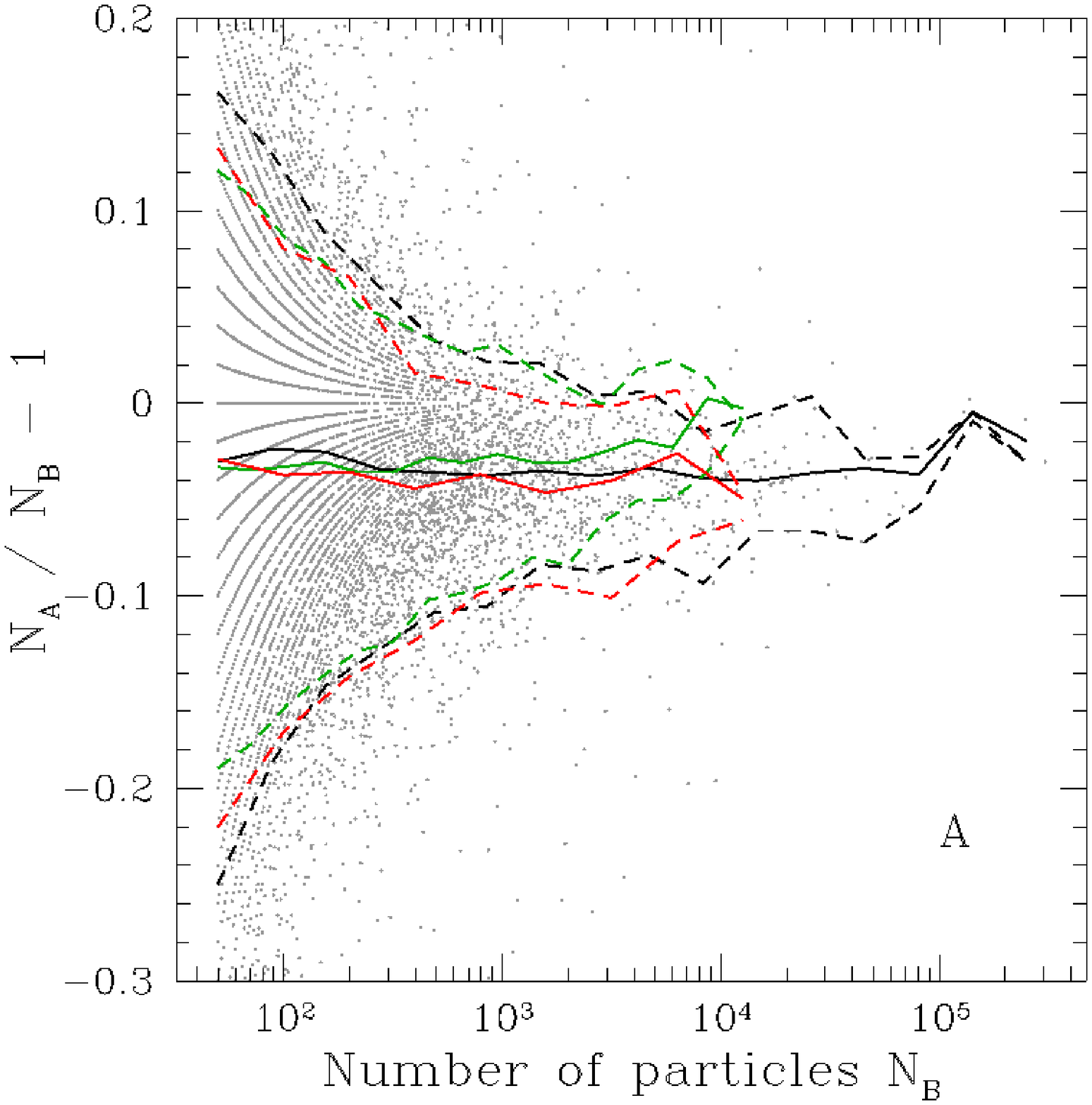,width=4.5cm}}
\put( 75,130){\psfig{file=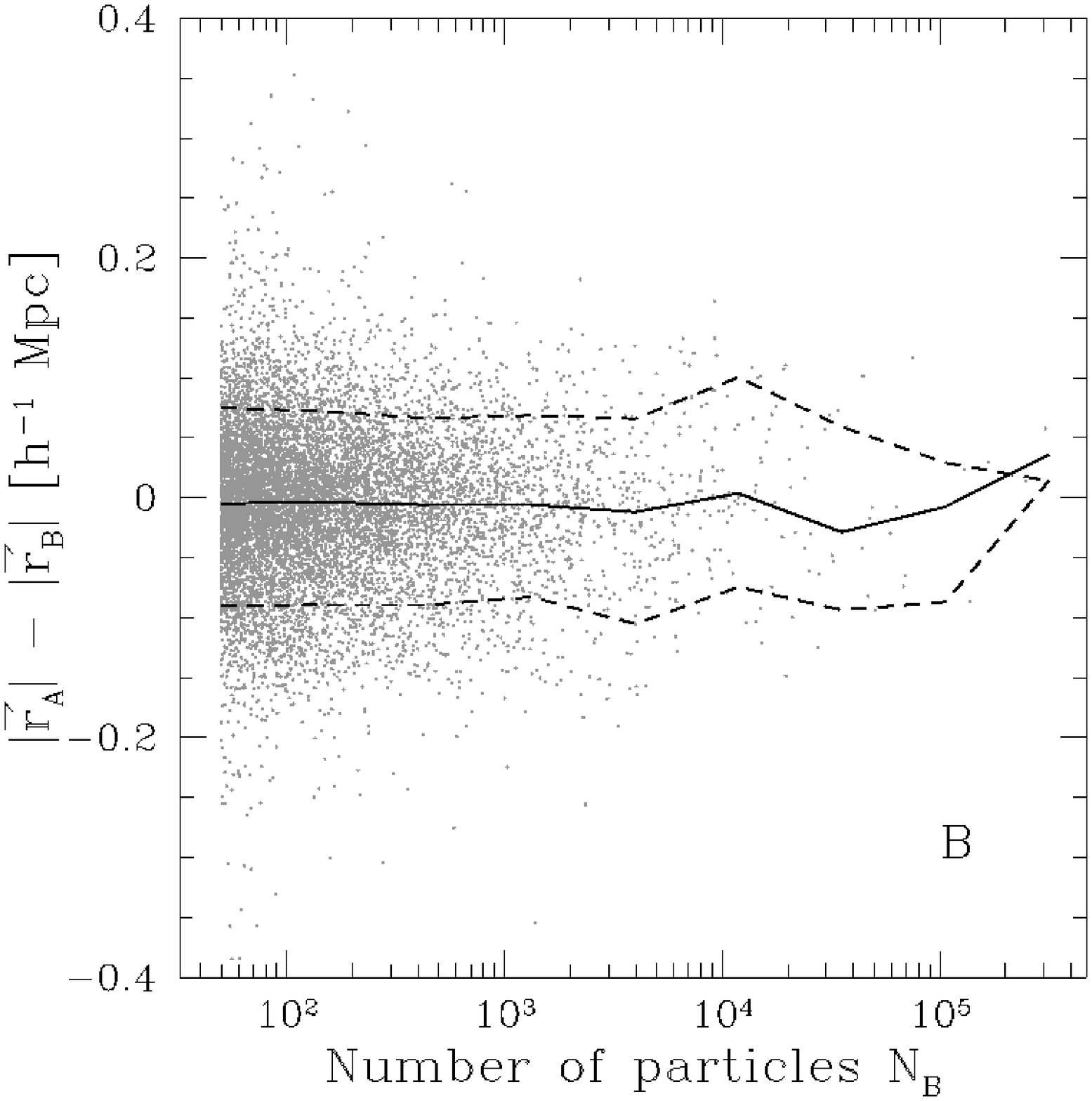,width=4.5cm}}
\put(200,130){\psfig{file=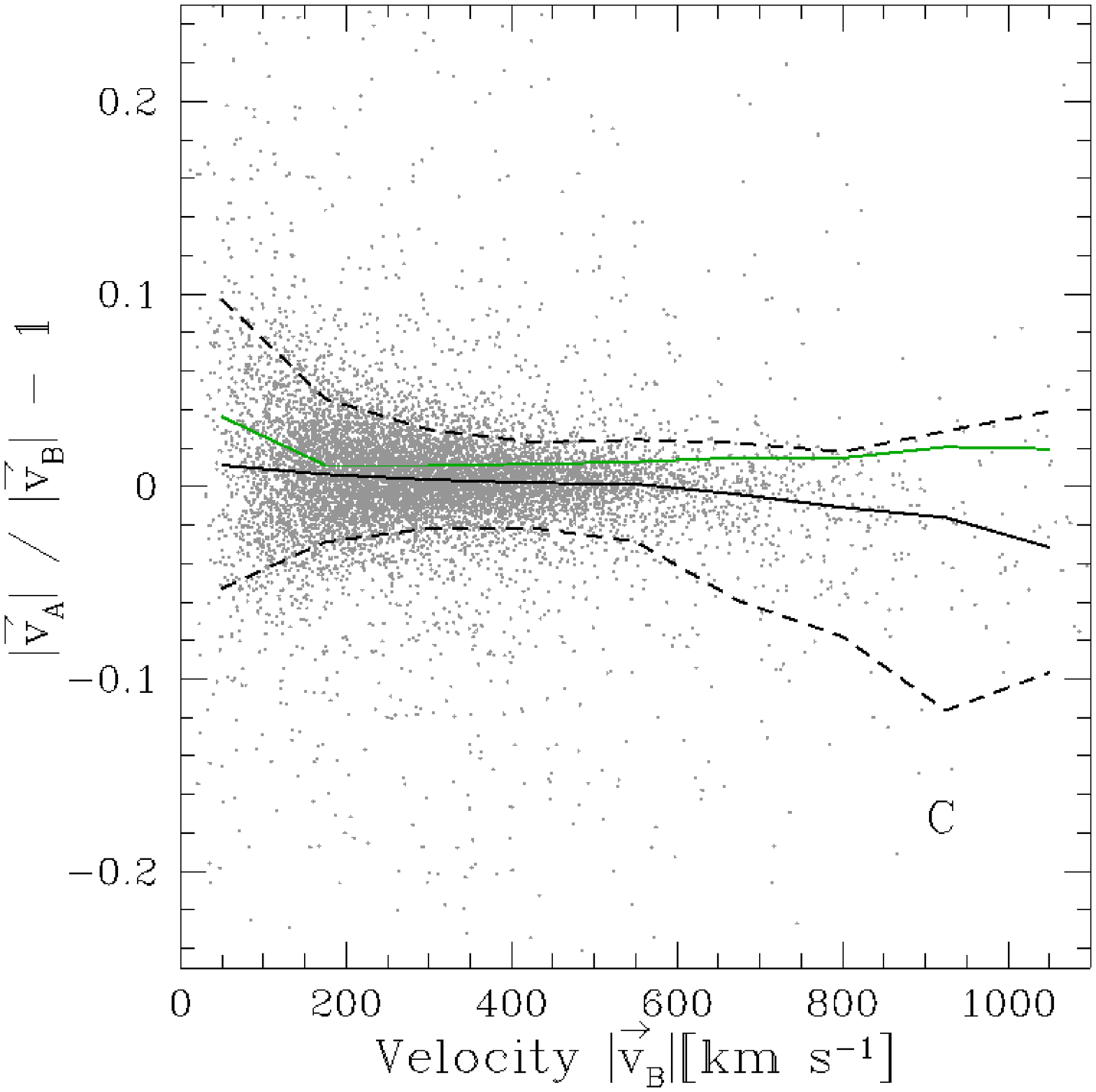,width=4.5cm}}
\put(325,130){\psfig{file=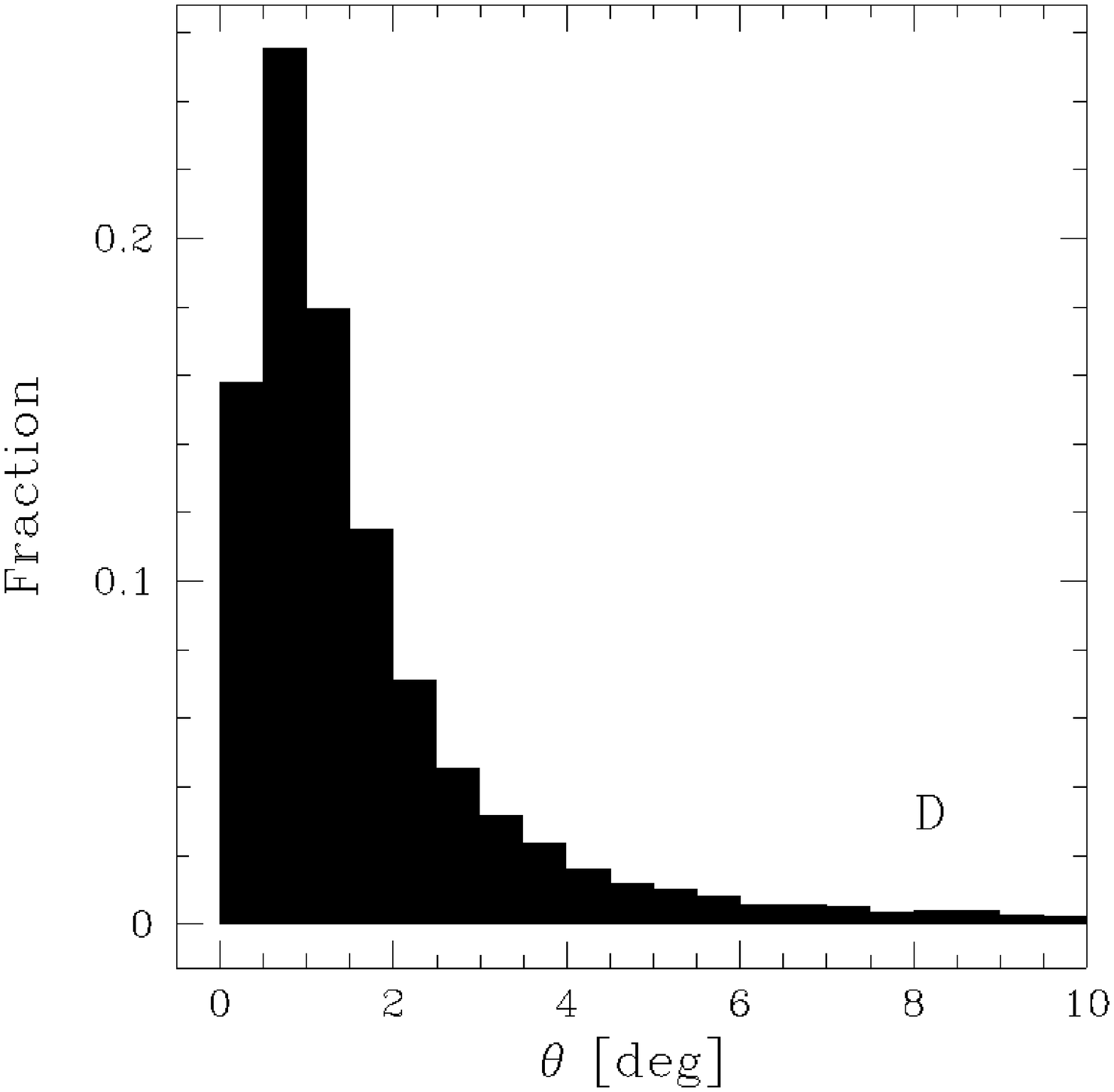,width=4.5cm}}
\put(-50,  0){\psfig{file=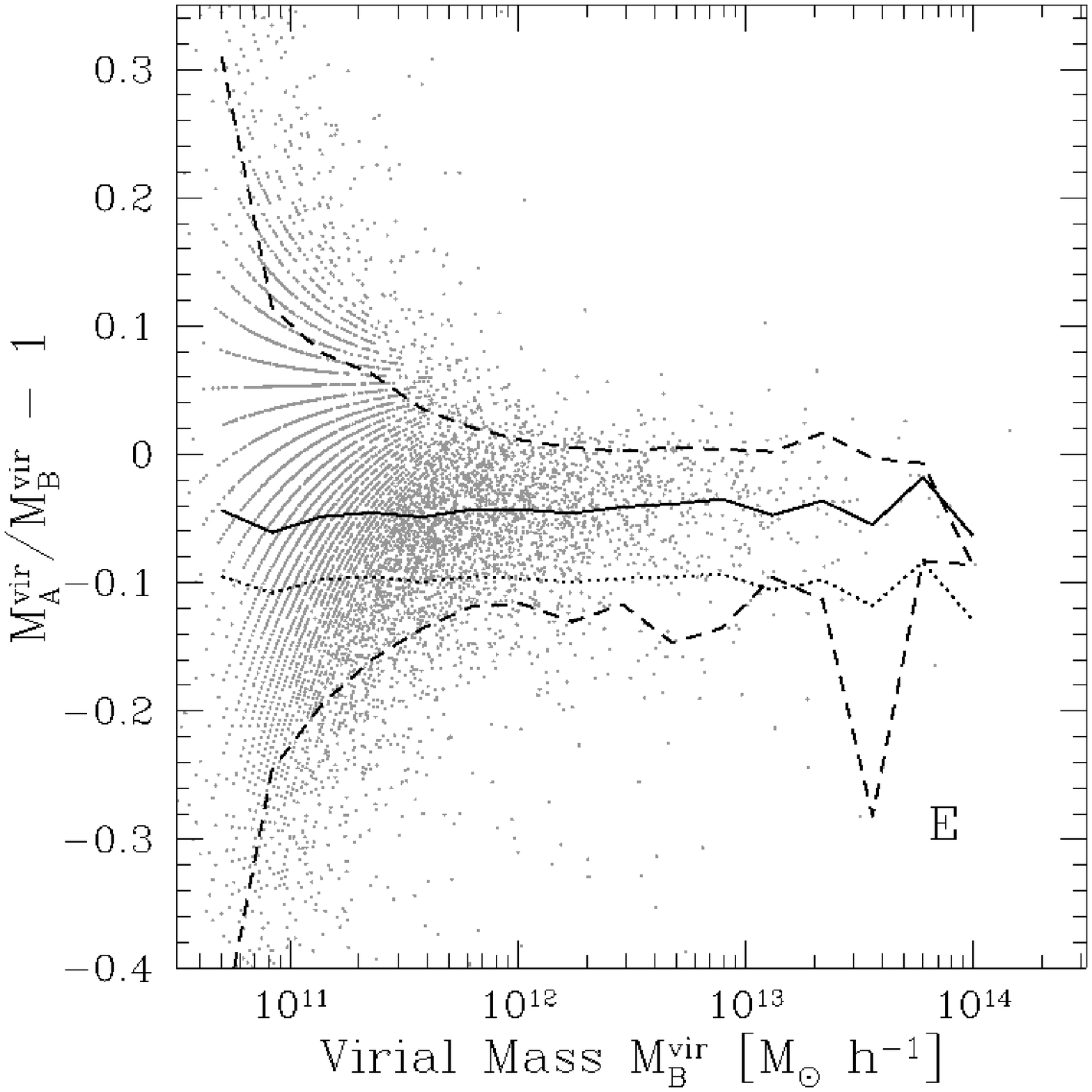,width=4.5cm}}
\put(75,   0){\psfig{file=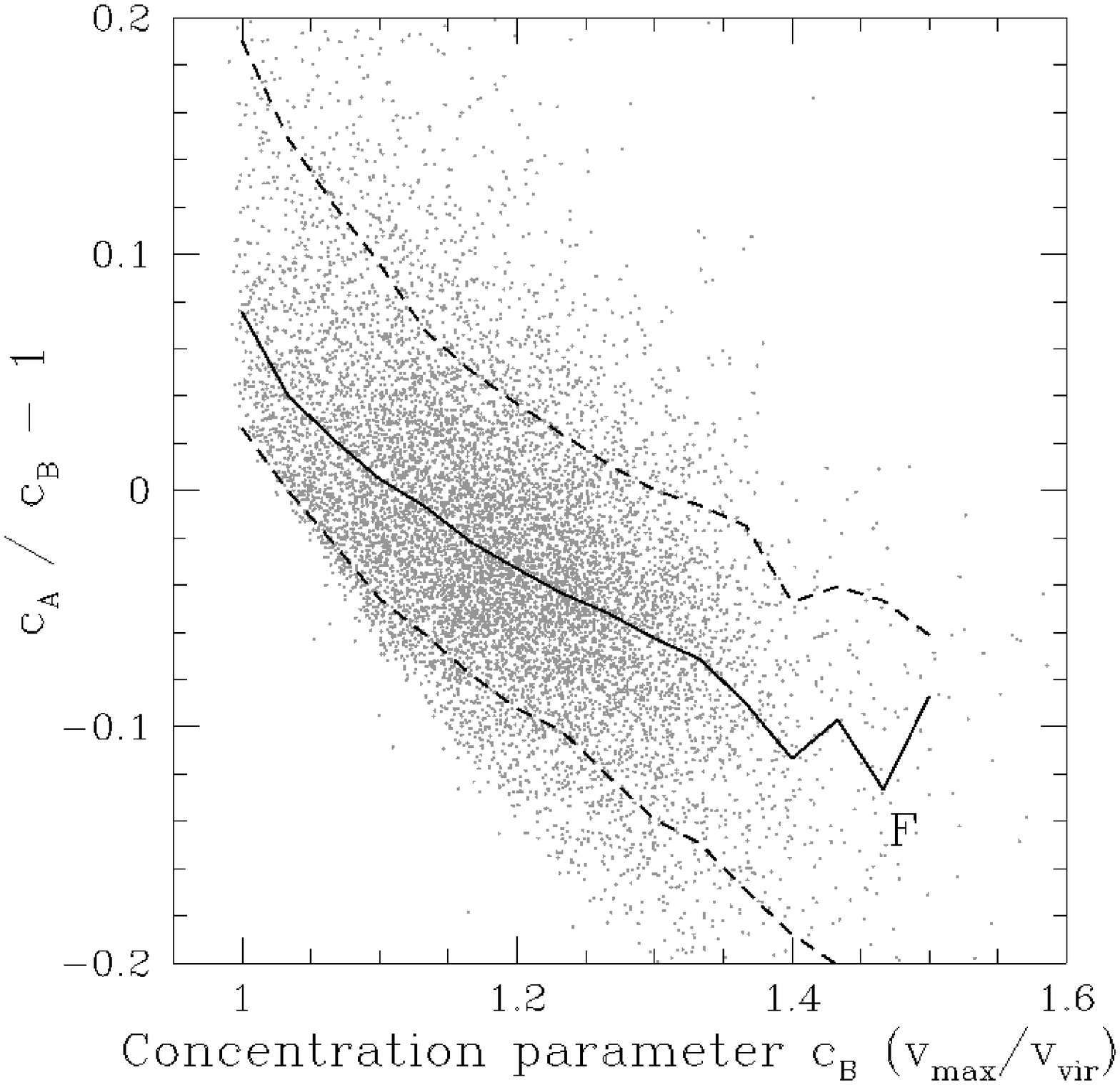,width=4.5cm}}
\put(200,  0){\psfig{file=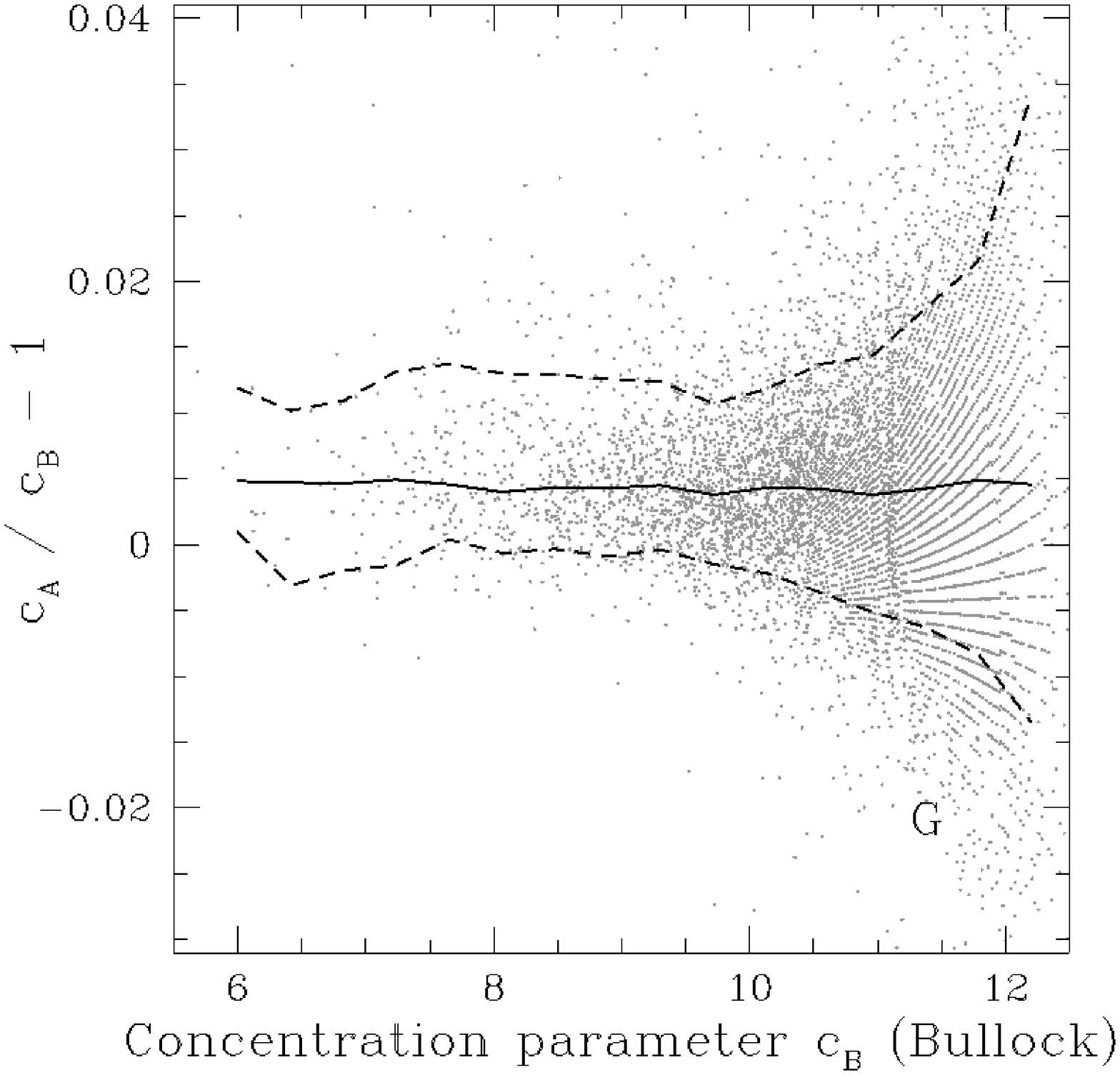,width=4.5cm}}
\put(325,  0){\psfig{file=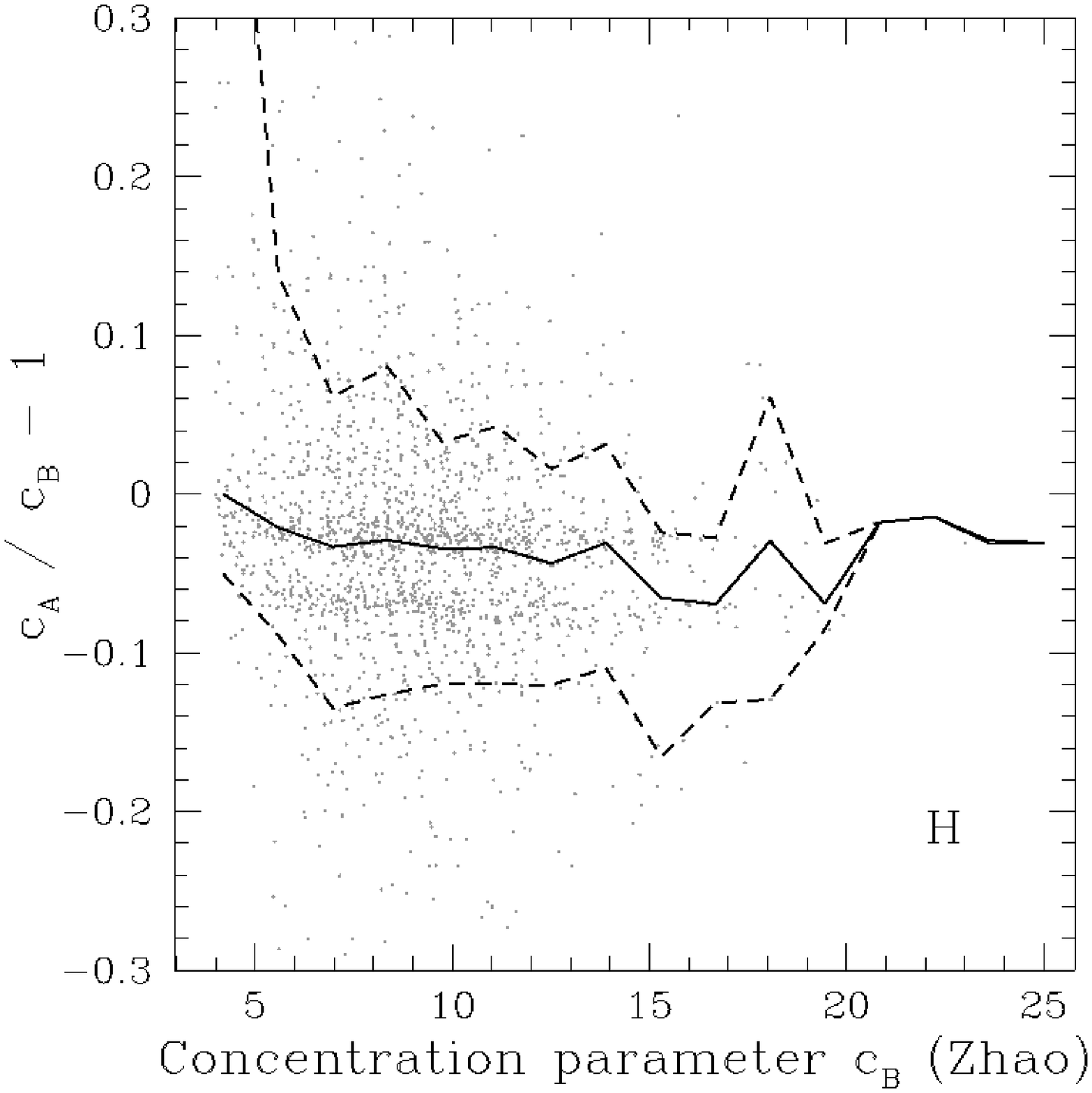,width=4.5cm}}
\end{picture}
\caption{Comparison of individual properties between rescaled and reference haloes. The top 
boxes show the relative difference in number of particles (A), the difference between halo positions 
as a function of the number of particles per halo (B), the relative difference between the modulus 
of the velocity vectors as a function of the peculiar velocity of haloes (C) and the histogram of 
the angle subtended by the velocity vectors of matched haloes (D). The bottom boxes show the virial 
mass (E) and three different halo concentration parameters, obtained from the ratio $v_{max}/v_{vir}$ 
(F), computed using the \citet{bullock01} recipe (G) and usig the \citet{zhao08} prescription (H). In 
all panels we show the full population of haloes. The black solid lines show the median, and black 
ashed lines enclose $80$ percent of the haloes. We also show the recovered number of particles for 
the simulation with a large box and with a lower resolution (panel A, green and red lines respectively), 
the median velocity offset for the large simulation (panel C, green solid line) and the median of the 
relative error in the virial masses before the correction of $\delta_{vir}(z,\Omega_m)$ via a NFW 
profile (panel E, dotted line.)}
\label{fig:indiv}
\end{figure*}

\begin{small}
\begin{table}
\caption{Rescaling parameters.  In the first column we show the simulation name, the second and 
third contain the scaling factor $s$ and the final redshift $z_A^f$, respectively, obtained after 
minimising Eq. \ref{eq:drms}. The last column shows the minimum $\delta_{rms}^2$ resulting from 
minimising the difference between the rescaled and desired linear fluctuation amplitudes. With the 
exception of simulation $Bbig$ where the range $[M_1,M_2]=[1\times10^{11},5\times10^{15}]$$h^{-1}$M$_\odot$ 
is used, in all simulations we use the mass range $[M_1,M_2]=[1\times10^{10},5\times10^{14}]$$h^{-1}$M$_\odot$ 
to perform the minimisation.}
\begin{center}
\begin{tabular}{cccc}
\hline
\hline
 Name & $s$ & $z_A^f$ & $\delta_{rms}^2$\\
\hline
$B$    & $1.128$ & $0.361$ & $2.6\times10^{-5}$ \\
$Blow$ & $1.128$ & $0.361$ & $2.6\times10^{-5}$ \\
$Bbig$ & $1.091$ & $0.325$ & $2.6\times10^{-5}$ \\
\hline
$Bo_{+2}$  & $0.779$ & $-0.383$ & $3.4\times10^{-7}$ \\
$Bo_{+1}$  & $0.876$ & $-0.191$ & $3.9\times10^{-7}$ \\
$Bo_{-1}$  & $1.164$ & $ 0.201$ & $9.5\times10^{-8}$ \\
$Bo_{-2}$  & $1.396$ & $ 0.425$ & $5.2\times10^{-7}$ \\
$Bs_{+2}$  & $1.000$ & $-0.147$ & $1.6\times10^{-7}$ \\
$Bs_{+1}$  & $1.000$ & $-0.077$ & $2.7\times10^{-7}$ \\
$Bs_{-1}$  & $1.000$ & $ 0.076$ & $5.3\times10^{-8}$ \\
$Bs_{-2}$  & $1.000$ & $ 0.153$ & $1.5\times10^{-7}$ \\
$Bo_{+1}s_{+1}$ & $0.876$ & $-0.273$ & $7.2\times10^{-8}$ \\

\hline
\hline
\end{tabular}
\end{center}
\label{tab:scaling}
\end{table} 
\end{small}

%%%%%%%%%%%%%%%%%%%%%%%%%%%%%%%%%%%%%%%%%%%%%%%%%%%%%%%%%%%%%%%%%%%%%%%%%%%%%%%%%%%%%%%%%%%%%%%%%

\subsection{$N$-body simulations}
\label{sec:simulations}

In order to test the scaling of haloes and their histories we use two sets of simulations. In a 
first set we use 6 simulations. The two main simulations are simulation $A$ which has a Millennium-like 
cosmology and simulation $B$ which has the background cosmology of the Bolshoi simulation. Both 
simulations contain $256^3$ particles in cubic volumes of $60$ and $67.68$h$^{-1}$Mpc of side length, 
respectively, which results in a particle mass comparable to that of the Millennium Simulation, 
$M_{p}\simeq10^{9}$h$^{-1}M_{\odot}$. This set of simulations also includes a low resolution version 
of simulations $A$ and $B$, with $128^3$ particles, and versions with volumes $\sim 100$ times bigger 
(with respect to $A$ and $B$), which we will refer to as $Alow$, $Blow$ and $Abig$, $Bbig$, respectively.
With the low resolution simulations we will minimise the rescaling function using the same parameters 
as for the $A$ and $B$ simulations; in this way we will only test the effect from the coarser resolution. 
In the case of the larger volume, the minimisation will be performed on a range of masses one order 
of magnitude higher than for $A$ and $B$ and will therefore include the effects from considering 
different scales in the power spectra, as well as a lower resolution.

Unless otherwise stated, the $B$ cosmology will be taken as the reference model and the simulations 
with the $A$ cosmology will be rescaled to resemble the former as closely as possible. 

The second set contains 9 low resolution simulations ($128^3$ particles) designed to measure the 
variation in the precision of the rescaling of the halo catalogues as the desired cosmology moves 
further away in the plane $\Omega_m - \sigma_8$ from the original parameters. In order to do that, 
we fix all parameters and vary only $\Omega_m$ (or $\sigma_8$) in $\pm 1\sigma$ and $\pm 2\sigma$ 
of its original value using $\sigma \sim 0.03$, which is in concordance with the standard WMAP7 
deviation for both parameters \citep{jarosik10}, giving us $8$ simulations. The remaining simulation 
was run varying both the$\Omega_m$ and $\sigma_8$ parameters by $+1\sigma$. The relevant parameters 
for both sets of simulations are detailed in Table \ref{tab:params}.  The range of masses over which 
the minimisation is performed is the same as for the $A$ and $B$ simulations, in order to mimic the 
accuracy of this procedure for a simulation with the particle mass resolution of the Millennium Simulation.

All the simulations were evolved from their intial redshifts using the public version of {\small GADGET2} 
\citep{gadget2}. The initial conditions were constructed using the public code {\small GRAFIC2} 
\citep{grafic2} and use exactly the same random seed in all cases. The halo catalogues, including 
substructure identification and merger histories, were constructed using the {\small SUBFIND} algorithm 
explained in detail in \citet{subfind}. The linking length parameter used is equal to $0.17$ times the 
mean interparticle separation and considering only groups with at least 10 particles. The simulations outputs 
consist of 100 steps equally spaced in $\log(a)$ between $z=20$ to $z=0$.

In the process of obtaining the $B$ simulations we search for the scaling factor $s$ and final redshift 
$z_A^f$ that provide the best fit between the actual linear fluctuation amplitudes via the minimisation 
of Eq (\ref{eq:drms}). These are shown in Table \ref{tab:scaling} for each simulation (see the Table 
caption for the mass ranges used in the minimisation). The table also shows the minimum value of 
$\delta_{rms}^2$ obtained.

%%%%%%%%%%%%%%%%%%%%%%%%%%%%%%%%%%%%%%%%%%%%%%%%%%%%%%%%%%%%%%%%%%%%%%%%%%%%%%%%%%%%%%%%%%%%%%%%%

\subsection{When is it acceptable to use the reduced AW10 method?}

In this subsection we perform a test that will allow us to infer in which cases it is acceptable
to ignore the quasi-linear correction of long wavelength contributions. This test consists on 
implementing the reduced and full algorithm and studying the resulting differences as a function 
of the boxsize of the simulation, and of the redshift. The latter is done since $k_{nl}(z)$ depends 
on the redshift. 

We use two sets of four $256^3$ particle simulations specific for this subsection with reference 
simulation boxsizes $L_{box} =$ 50, 100, 500 and 1000 $h^{-1}$Mpc. The first set has the same 
cosmological parameters than simulation $A$ and the second one has the parameter set of simulation 
$B$. We will analyse the $z=$0, 0.5, 1, 2 and 4 outputs.

The top panel of Figure \ref{fig:sigmas} shows the 1D rms difference in the particle positions, for 
the case where the AW10 method is applied in full and in its reduced version (solid and dashed lines,
respectively); each line shows the change in the error as the simulation box increases to the right.  
As can be seen, for simulations of $\sim 50 h^{-1}$Mpc a side, the reduced rescaling algorithm 
produces positions that are as precise as those obtained using the full method.

The bottom panel shows the ratio between the 1D rms differences obtained from the reduced and 
full AW10 methods for different values of the simulation box, as the redshift increases to the right. 
In this case it can be clearly seen that the small simulation box (with side $\sim 2\pi/k_{nl}(z=0)$) 
shows almost a unit ratio for all the redshifts explored.

The fact that the large-scale correction is not important in simulations of $\sim 50 h^{-1}$Mpc 
a side is expected since, by construction, the displacement fields are smooth on modes larger than 
$k_{nl}$. Note that this correction depends on the cosmologies selected for the simulations, and 
therefore the results shown in this subsection are only presented as a qualitative example.

%%%%%%%%%%%%%%%%%%%%%%%%%%%%%%%%%%%%%%%%%%%%%%%%%%%%%%%%%%%%%%%%%%%%%%%%%%%%%%%%%%%%%%%%%%%%%%%%%
%%%%%%%%%%%%%%%%%%%%%%%%%%%%%%%%%%%%%%%%%%%%%%%%%%%%%%%%%%%%%%%%%%%%%%%%%%%%%%%%%%%%%%%%%%%%%%%%%

\section{Results}
\label{sec:results}

\label{ssec:haloprops}

\begin{figure*}
\begin{picture}(400,340)
\put(-50,-20){\psfig{file=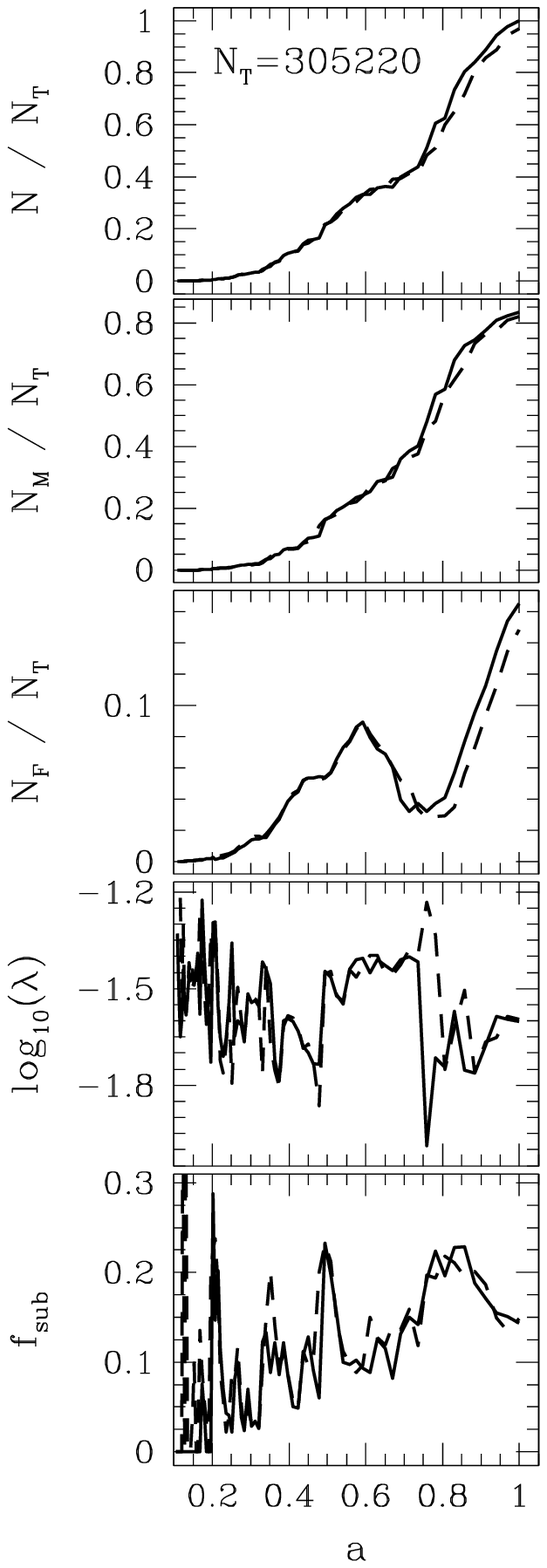,width=5.cm}}
\put(75, -20){\psfig{file=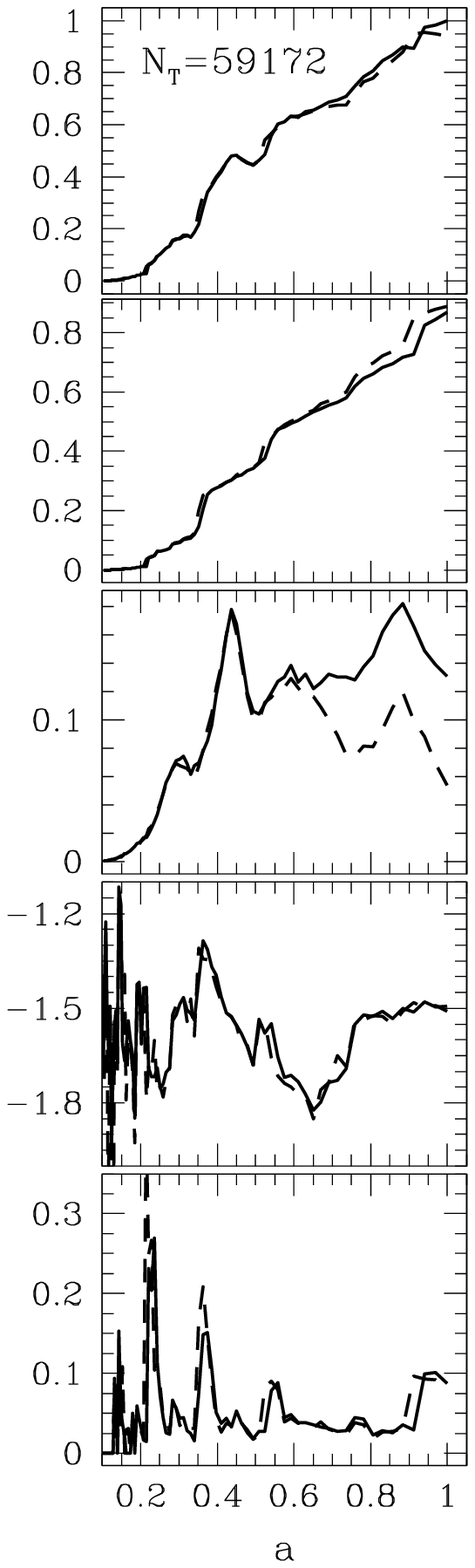,width=5.cm}}
\put(200,-20){\psfig{file=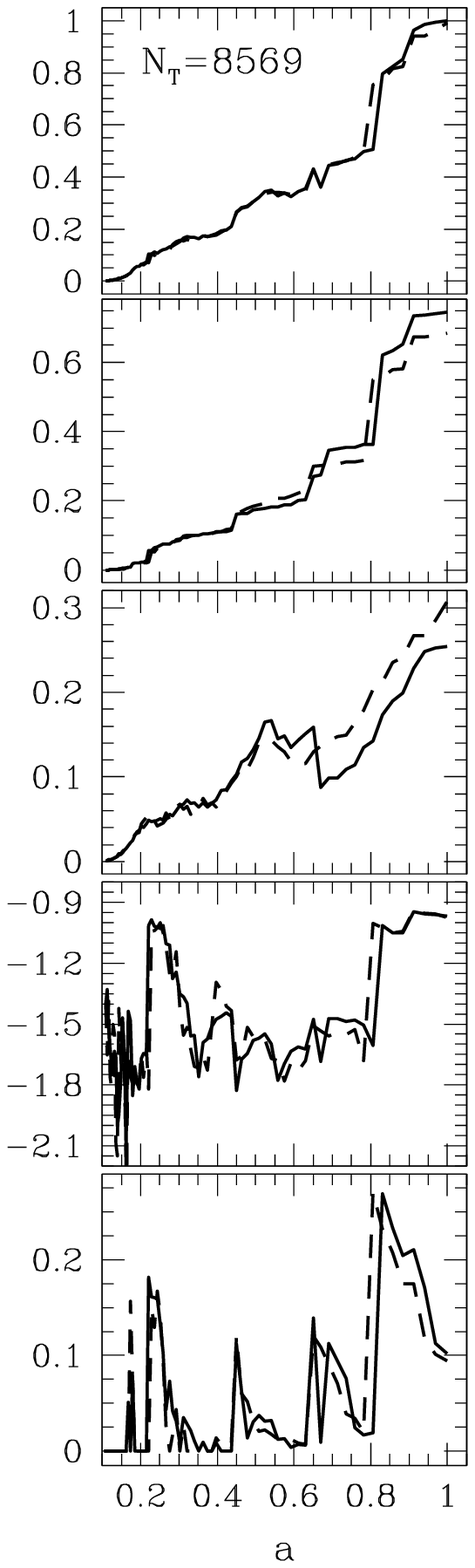,width=5.cm}}
\put(325,-20){\psfig{file=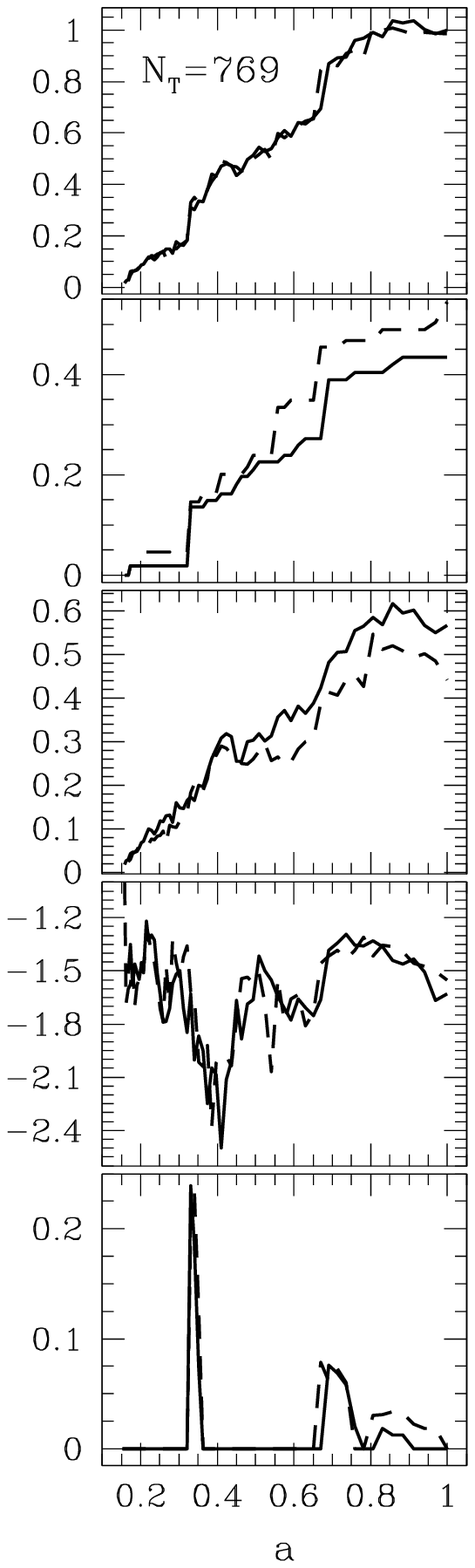,width=5.cm}}
\end{picture}
\caption{Comparisons between individual haloes belonging to the rescaled and reference catalogues. The 
number of particles per halo is shown in the upper panels. From top to bottom the panels show the cumulative 
accretion history, cumulative accretion via mergers, cumulative accretion of individual particles, the evolution 
of the spin parameter and the fraction of mass in substructures, all as a function of the expansion parameter 
$a$.  The solid lines show the true evolution; the dashed lines are for the evolution in the rescaled catalogue. 
The haloes correspond to the 1st, 10th, 100th, and 1000th largest haloes in simulation $B$ and their FoF 
masses are $4.2\times10^{14}$, $8.2\times10^{13}$, $1.2\times10^{13}$ and $1.1\times10^{12}$ $h^{-1}$M$_\odot$ 
respectively.}
\label{fig:histories}
\end{figure*}

In this section we apply the rescaling to the simulations presented above, and perform tests on the
recovery of the properties of individual haloes, including their detailed growth histories.  We also study
the recovery of statistical properties of the global population of haloes. From this point on we apply
the reduced version of the AW10 algorithm to dark matter haloes in all the simulations except $Bbig$,
to which the full method is applied since has a boxsize for which the quasi-linear correction is important.

%%%%%%%%%%%%%%%%%%%%%%%%%%%%%%%%%%%%%%%%%%%%%%%%%%%%%%%%%%%%%%%%%%%%%%%%%%%%%%%%%%%%%%%%%%%%%%%%%

\subsection{Comparison of Individual Halo Properties}

We compare the properties of haloes of at least $50$ particles in the numerical simulations run with the 
desired cosmological parameters ($B$ simulations, the reference model) to the haloes from simulations $A$ 
(with a different cosmology) rescaled to the cosmology of the $B$ simulations (rescaled haloes). These 
haloes are allowed any number of member particles (at least $10$).
In order to do this we match haloes that share the largest percentage of particles via their {\small GADGET2} 
identifying particle ID (referred to as matched haloes); this can be done as the initial conditions are 
constructed using the same random seed. The percentage of haloes in the $B$ simulations that have a 
matched halo in the $A$ simulation is higher than $99\%$ in all cases.

We compare the properties of matched haloes in Figure \ref{fig:indiv}. The number of particles of the 
recovered haloes shows a slight underestimation of $<5$ percent, and a clear increase in the dispersion 
that decreases with halo mass and becomes $10$ percent for haloes of $\sim 200 $ particles. Results for 
a lower resolution simulation ($Blow$) in green, and for the simulation with $\sim 100$ times larger 
volume ($Bbig$) in red are also shown. As can be seen, neither the lower resolution nor the larger 
amplitude over which the minimisation is done affect the precision of the recovered number of particles 
per halo for haloes with $>200$ particles. The catalogue of simulation $Bbig$ shows a slightly broader 
range of recovered number of particles for haloes with less particles, but it is not clear whether this 
is a result of noise or the larger box of the simulation. The results for the $Blow$ and $Bbig$ 
simulations are also similar to those of the $B$ simulation in most of the comparisons shown in the other 
panels of this figure. Therefore, in order to improve clarity, in the remaining panels we will only show
the results for the $B$ simulation except for the cases where there are noticeable differences with 
resolution or box size.

As can be seen the positions of haloes are well recovered to a precision of $0.1h^{-1}$Mpc for the 
catalogue of simulation $A$ (difference between percentiles $10$ and $90$ and the median). The velocities 
show relative differences of less than $\sim 5$ percent, and small changes in the direction of the velocity 
vector with a mode of $1$ degree. As can be seen in the figure, the  velocities tend to be biased high for 
$v<350$km/s, and biased low for $v>700$km/s, although always below a 5 percent difference. The precision 
in the recovered halo positions and velocities are consistent with those reported by AW10 in their Figure 
8 using the full implementation of the algorithm.

The larger box (blue line) produces an increase in the bias at low peculiar velocities but removes it at 
the large velocity end. Also, the amplitude of the differences between velocities increases more rapidly 
for low velocities in the catalogue of the large volume simulation.

We compute the virial mass of haloes as the mass inside a sphere that contains an average virial 
overdensity of $\delta_{vir}(z,\Omega_m) \approx 18\pi^2 (1+0.4093x^{2.7152})(1+x^3)^{-1}$, where 
$x = (1/\Omega_m(z=0) - 1)^{1/3}(1+z)^{-1}$ \citep{nakamura}. Since the methodology we apply does not 
involve re-identifying the haloes using dark-matter particles, the virial mass for the rescaled haloes 
resulting from applying Equation (\ref{eq:mass_cor}) is underestimated by a $\sim 10$ percent 
(panel E, dotted line). However, by considering a correction due to the difference in $\delta_{vir}(z,\Omega_m)$ 
which is $15$ percent higher for the cosmology of the recovered haloes, we are able to reduce 
the discrepancy in the virial masses to below $5$ percent. To do this correction we assume a 
NFW dark matter density profile \citep{nfw} for each halo, compute the integral of the 
assumed density profile above the corresponding virial overdensity in each cosmology using a 
concentration parameter given by \citet{bullock01}, and then multiply the rescaled mass by the 
ratio between the integrated mass in the target and base models.

We also explored the effect of adopting different concentrations on the recovered virial masses 
(Figure \ref{fig:indiv}). We used three different concentrations, (i) the proxy $v_{max}/v_{vir}$ 
(where $v_{max}$ is the maximum circular velocity of the halo and $v_{vir}$ is the circular velocity 
at the virial radius), (ii) the recipe given by \citet{bullock01}, and (iii) from the fit provided by 
\citet{zhao08}. Our results indicate a negligible effect on the mass, which neither improves nor 
diminishes the level of agreement between rescaled and reference masses, independently of the 
definition of concentration applied.

Regarding the recovery of the concentrations of the haloes in the target cosmology, the individual 
concentrations defined as $v_{max}/v_{vir}$ are affected by a bias of $\sim 10$ percent with a 
strong dependence with the concentration in the reference cosmology.  On the other hand, the 
concentrations computed using the mass dependence reported by \citet{bullock01} show a negligible 
difference between rescaled and reference cosmologies, with no dependence on the halo virial mass.
The \citet{zhao08} concentrations are obtained using their dependence on the cosmological time 
when the main progenitor acquired $4$ percent of the final FoF halo mass. This quantity can be 
computed using the information in the halo merger trees; however, due to the numerical resolution 
of our simulation, this quantity can only be calculated for the $\sim 2000$ most massive haloes. 
For these haloes, the concentration shows no offset between the rescaled and target simulations 
but its dispersion is larger than for the \citet{bullock01} concentrations, particularly for 
low $c$ values.

Even though the recovery of individual properties shows some biases and non-negligible scatter, 
the quantification of these effects can be used to gauge what studies can be performed with rescaled 
simulations and the statistics that are expected to be affected by these uncertainties.

\begin{figure}
\includegraphics[scale=0.4]{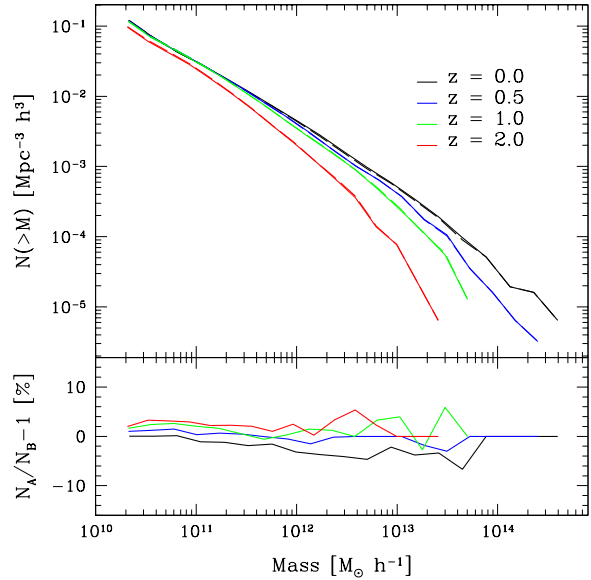}
\caption{Mass function for rescaled and reference FoF halo catalogues (solid and dashed lines, 
respectively) for different redshifts (different colours as indicated in the figure key). 
The lower sub-panel shows the relative difference between the rescaled and reference catalogues, 
in the same colour scheme.}
\label{fig:mf}
\end{figure}

\begin{figure}
\centering
\includegraphics[scale=0.4]{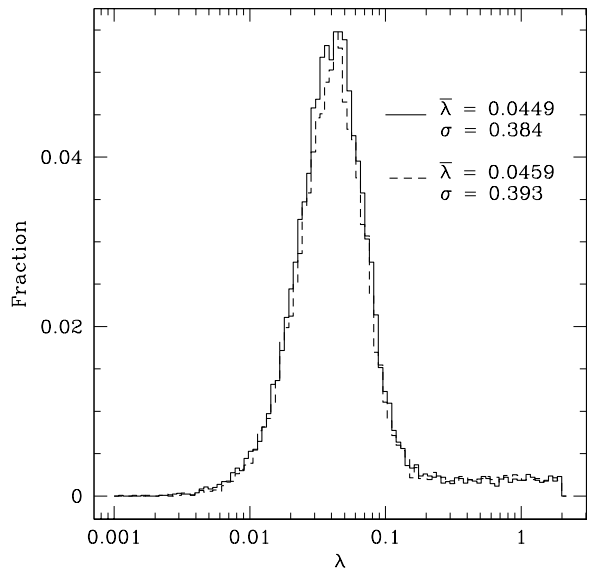}
\caption{Distribution of halo spin parameters for the rescaled (dashed lines) and reference
(solid) catalogues. The mean value and standard deviation of the log-normal distributions are 
shown in the legend.}
\label{fig:lam}
\end{figure}

Our findings on the accuracy of the properties of haloes are consistent with those found by AW10,
which indicates that applying the full method to particles or haloes produces only small differences.
We confirm this estimate
using our simulations, where we only find negligible differences in the resulting accuracy when using
individual particles.

%%%%%%%%%%%%%%%%%%%%%%%%%%%%%%%%%%%%%%%%%%%%%%%%%%%%%%%%%%%%%%%%%%%%%%%%%%%%%%%%%%%%%%%%%%%%%%%%%

\subsection{Examples of halo evolution}

For a more detailed comparison between the rescaled and reference halo catalogues, we also compare 
the detailed histories of individual friends-of-friends (FoF) haloes.
  
We choose the first, 10th, 100th and 1000th most massive haloes and show them in Figure \ref{fig:histories}. 
The panels show from top to bottom, the accretion history, the accretion via mergers and via smooth, 
individual particle infall, the evolution of the dimensionless spin parameter, and the fraction of mass
in substructures.   The number of particles per halo decreases to the right, and range from FoF masses 
of $4.2\times10^{14}$h$^{-1}M_{\odot}$ to $1.1\times10^{12}$h$^{-1}M_{\odot}$.  Regardless of the halo 
mass, the individual accretion histories, spin parameter evolution, and fraction of mass in substructures, 
are remarkably well recovered.  The only noticeable differences are shown in the mass accretion via mergers 
and individual infall of particles, which cancel out as when one is overestimated the other compensates, 
showing the effects from confusion between accretion of small mass haloes and infalling field particles. 
This effect can be appreciated more clearly in the mass accretion history of the 10th largest halo in 
Figure \ref{fig:histories}.

These results indicate that individual merger trees should be reasonably suitable for rescaling and later 
use for semi-analytic type galaxy formation modeling. The history of mass accretion via mergers would ensure 
a $10$ percent accuracy in the population of satellite galaxies, whose added stellar masses should be even 
more precise (according to the well recovered fraction of mass in substructures), and the accurate evolution 
of the spin parameter should ensure reasonable estimates of galaxy disc sizes.

%%%%%%%%%%%%%%%%%%%%%%%%%%%%%%%%%%%%%%%%%%%%%%%%%%%%%%%%%%%%%%%%%%%%%%%%%%%%%%%%%%%%%%%%%%%%%%%%%

\subsection{Statistical properties of haloes and their growth histories}

If the rescaled simulation is used to produce statistics of large populations, either of dark-matter 
haloes or simulated galaxies obtained via semi-analytic techniques, it is necessary to estimate how 
they are affected by the rescaling process.

We first study the mass function of dark-matter haloes.  Figure \ref{fig:mf} compares the mass functions 
of the rescaled and reference haloes (dashed and solid lines, respectively) for different redshifts 
from $z=0$ to $z=2$. The agreement is excellent with only minor differences. The relative differences 
shown in the lower panel indicate that there is a mild, although clear trend that starts as an 
underestimation of abundances at $z=0$ but then tends to overestimate them by larger amounts at higher 
redshifts. At low redshift the rescaled catalogue shows a slight underestimation of the number density 
of haloes which is very mild at low masses $M\sim10^{10}$h$^{-1}M_{\odot}$ at the $1$ percent level, 
but increases to a $5$ percent lower abundances at $M\sim10^{13.5}$h$^{-1}M_{\odot}$.  The highest 
redshift shown in the figure indicates a flat overestimation of abundances of a $\sim 3-5$ percent.

These effects are small in comparison to the current precision of measurements of the mass function in 
clusters of galaxies \citep{gladders07} and galaxy luminosity functions (for example in the SDSS, 
see \citealt{blanton03}) which, in simulations, are highly influenced by the underlying mass function of 
dark-matter haloes (see for instance \citealt{cole00}).  However, the determination of a power spectrum 
from a mass function requires high precision measurements \citep{sanchez02} and the accuracy of the 
rescaling technique could be important in this case.

The spin parameter influences the resulting properties of disc galaxies in semi-analytic models 
\citep{cole00, lagos09, tecce10}. Therefore it is important to check whether the rescaling of a 
numerical simulation could produce biased distributions of spin parameters. Figure \ref{fig:lam} 
shows these distributions for the rescaled and reference catalogues (dashed and solid lines, respectively), 
and as can be seen, the distributions are consistent with one another. The distributions of disc sizes 
would therefore be expected to be reliable in the rescaled simulation. This is also in agreement with 
the evolution of the spins of individual matched rescaled and reference haloes shown in Figure 
\ref{fig:histories}, which are consistent with one another at all times. 

\begin{figure*}
\includegraphics[scale=.6]{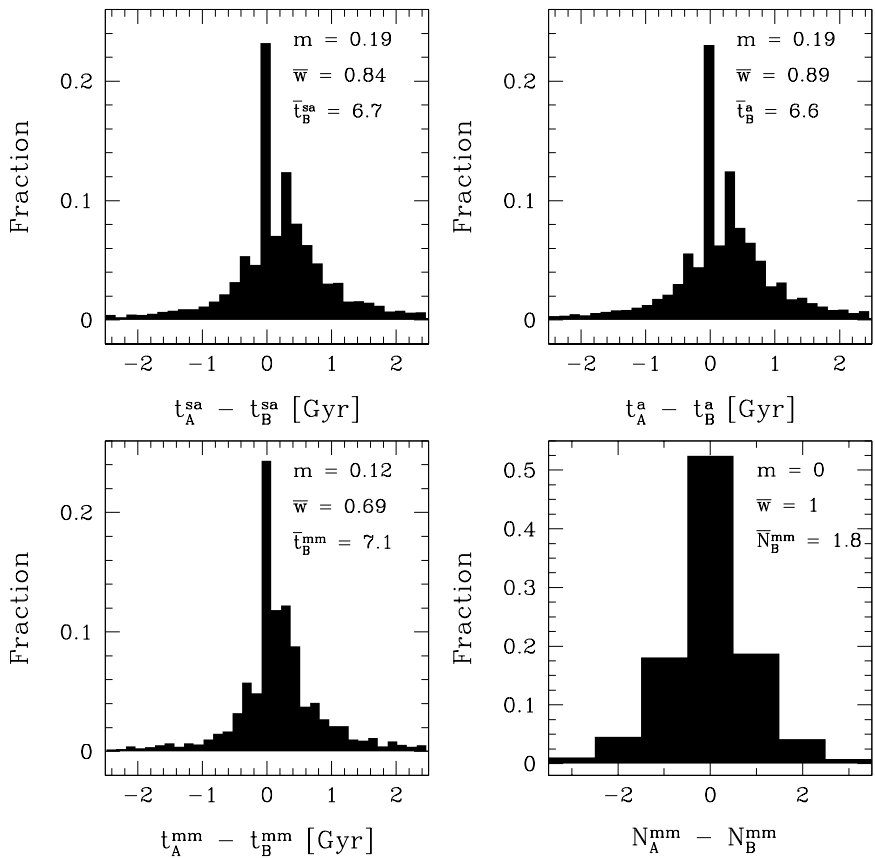}
\caption{Distributions of differences between merger-tree properties for dark-matter haloes in the 
rescaled and reference model. Top-left: distribution of the error in the time since the mass added 
over all the branches of the merger tree reached half the final mass. Top-right: same as top-left 
for the time since the halo in the main branch of the merger tree attained half its final mass. 
Bottom-left: same as previous panels for the time since the last major merger (with at least a $0.3$ 
mass ratio). Bottom-right: differences between the number of major mergers for haloes which have 
undergone at least one major merger. The median $m$ and mean width $\bar w$ of the distributions 
are given in the legend of each panel; we also show the average quantities in the reference cosmology. 
For the upper panels and the lower-left panel, the units are Gyr.}
\label{fig:times}
\end{figure*}

Given that the stellar population of a galaxy is highly dependent on the mass accretion history of a 
halo, either via mergers or smooth infall, a statistical study of the merger trees of haloes can 
also be used to test the adequacy of the rescaling.  Figure \ref{fig:times} shows the differences 
between rescaled and reference models for four particular characteristics of merger trees. The 
quantity shown on the top-left panel corresponds to the time when half the total mass of a halo was 
formed in the tree $t^{sa}$ (i.e. including all the mass in satellites that will later merge with 
the central halo); by subtracting any cooling and star formation timescales involved, this would 
correspond to a stellar age for the final galaxy. As can be seen, the differences show a clear peak 
at $0$Gyr and a median at $\sim 0.19$ Gyr, with a mean witdh of $0.84$Gyr.  
On a $z=0$ galaxy population the former difference would produce little effects on the resulting 
galaxies, but the latter may spuriously broaden distributions of galaxy properties such as colours.

The second statistics, the time since the halo in the main branch of the merger tree attained half 
of its final mass, $t^a$ (top-right panel of Figure \ref{fig:times}), would be related to the time 
of assembly of the mass of the final galaxy and shows an offset of $\sim 0.19$Gyr and a mean witdh 
of $0.86$Gyr. Similar results are obtained for the time since the last major merger $t^{mm}$ 
(bottom-left) which should correlate with the time since the last starburst (the actual starburst 
should be more recent than this quantity when taking into account the dynamical friction that affect 
galaxies), which shows a similar offset of $\sim 0.12$Gyr on average, and a similar distribution 
width of $0.69$Gyr.  

Since the timescales of last starburst, star formation, and stellar mass assembly are affected in 
a similar way, we do not expect important changes on the spectral shape of a galaxy obtained using 
rescaled merger-trees, with the exception of the effect from a general shift in age towards older 
populations by $\sim 0.2$Gyr. The lower-right panel of Figure \ref{fig:times} shows the histogram 
of the difference between the number of major mergers undergone by the rescaled and reference haloes 
between $z=4$ and $0$. As can be seen, the distribution is symmetric with a clear peak at zero 
(which contains more than $50$ percent of the sample), and a maximum difference of $2$ or more major 
mergers for about $10$ percent of the haloes.

AW10 presented comparisons between semi-analitic galaxies obtained from rescaled and direct 
simulations, showing an effect that may be the product of these offsets, the slightly fainter K-band 
magnitudes (by $0.1$ mag) in their rescaled simulations. 

\begin{figure}
\includegraphics[scale=0.4]{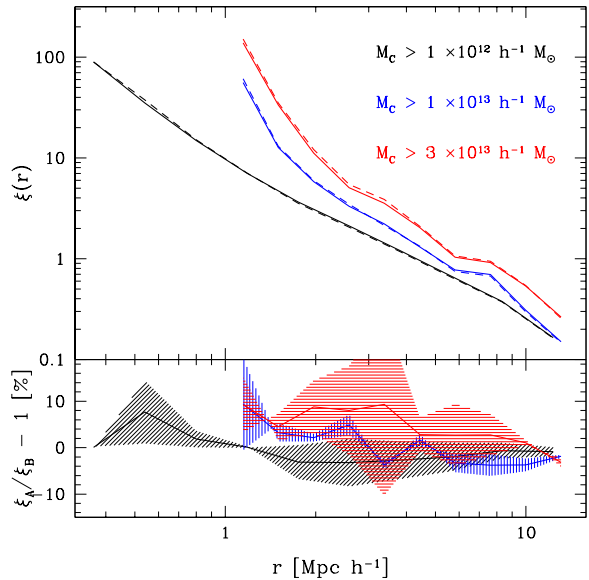}
\caption{Cross correlation functions between haloes in the rescaled $A$ (dashed lines) and 
reference $B$ (solid lines) catalogues with different lower limits on FoF mass (different colours 
shown in the key) and the full population of haloes (with at least $10$ particles). The lower 
sub-panel shows the relative differences between the correlation functions obtained from the 
rescaled and reference catalogues; the shaded regions show the uncertainties computed using 
the boostrap resampling technique.}
\label{fig:xi}
\end{figure}

The study of the effect of the rescaling of merger trees on the inferred positions of haloes performed in 
Section \ref{ssec:haloprops} showed that these are not badly affected by the process.  However, the 
correlation function could show changes due to the dispersion in the positions shown in the top-left 
panel of Figure \ref{fig:indiv}. Figure \ref{fig:xi} shows the resulting cross-correlation functions 
from the rescaled and reference haloes. For this test we choose cross-correlations over auto-correlation 
functions in order to increase the signal-to-noise ratio of our measurements (see for instance 
\citealt{bornan,lacerna}). Halos of different lower limits on FoF mass are used as centres, whereas all 
haloes in the simulations are used as tracers for these measurements. We choose three different lower 
limits on the mass of centre haloes ($M_C>10^{12}$h$^{-1}M_{\odot}$, $M_C>10^{13}$h$^{-1}M_{\odot}$ and 
$M_C>3\times10^{13}$h$^{-1}M_{\odot}$) such that they bracket the non-linear mass for the cosmology of 
simulation $B$, $M_{nl}\sim10^{13}$h$^{-1}M_{\odot}$, around which the bias factor shows a clear increase 
(for haloes with mass $M_{nl}$, the bias factor is $b=1$, for more details see for instance \citealt{sheth01}).

As can be seen, at scales $r>0.5$, $2.5$, and $4$h$^{-1}$Mpc for masses $M_C > 1 \times 10^{12}$, $1 
\times 10^{13}$ and $3 \times 10^{13}$h$^{-1}$M$_\odot$, respectively, the precision of the correlation 
function of the rescaled haloes is better than $5$ percent (see the ratios on the lower sub-panel). 
Given the offset in the number of particles between the reference and rescaled haloes, we also tested 
whether using the same equivalent lower limit in number of particles improved this comparison, but find 
very similar results. This is also the case for the effect of the dispersion between rescaled and reference 
halo masses (number of particles), since the resulting change is of only a few percent, which influences 
the clustering amplitude by factors below the offsets originating from the rescaling procedure.  

This level of precision is of the order of that obtained for large surveys at $z=0$ such as the SDSS 
(see for instance \citealt{zehavi04}). Furthermore, other effects such as the assembly bias 
\citep{gao05} are expected to produce variations on the clustering of haloes and, consequently, 
galaxies to a $10$ percent, which indicates that rescaled haloes could be used to test the 
detectability of this particular effect. However, \cite{Wu08} point out that the precision of future 
surveys such as the Dark Energy Survey (DES, \citealt{tucker07}) and the Large Synoptic Survey 
Telescope (LSST, \citealt{abell09}) will be much better than that in the SDSS, and could irequire
new runs with the desired cosmology rather than rescaled halo catalogues.

\begin{figure}
\centering
\includegraphics[scale=0.4]{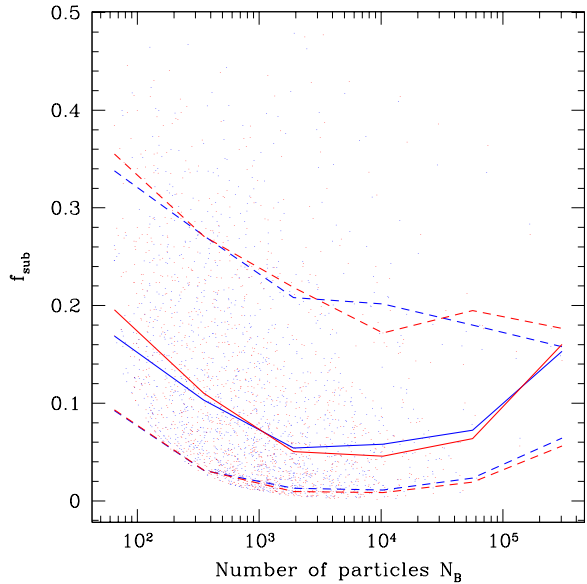}
\caption{Dependence of the median (solid lines), $10$ and $90$ percentiles (dashed lines)
of the difference between the recovered and true fraction of mass in substructures as a 
function of the number of particles per halo. Blue and red lines correspond to the rescaled 
and reference halo catalogues, respectively. The uprising of $f_{sub}$ at small masses 
may correspond to a resolution effect.}
\label{fig:fsub}
\end{figure}

As a final test, we study the changes introduced by the rescaling of haloes in the fraction of 
FoF particles in substructures (i.e., not restricting substructures to the sphere contained 
in one virial radius). Figure \ref{fig:fsub} shows this fraction as a function of the number of 
particles per halo, for the rescaled and reference halo catalogues (lines and dots in blue and red, 
respectively).  Both the median relation, and the envelope of $80$ percent of the haloes show an 
excellent agreement between the two catalogues, with differences that only amount to a $10$ percent. 
This statistics is related to different properties of the galaxies inhabiting these haloes. On the one 
hand, it would influence the variation of the amplitude of clustering as a function of subhalo mass 
(or galaxy luminosity), since the average host halo mass of a selection of subhaloes can be affected 
if this fraction is not accurately recovered. On the other hand, this result is in agreement with our 
earlier claim that we would not expect to find correlated changes in the spectra of semi-analytic 
galaxies obtained from the rescaled trees. This is due to the fact that the changes in the characteristic 
timescales of the merger trees are slightly biased but these biases go all in the same direction, 
otherwise we would expect to find shifts in the fraction of mass in subhaloes. This is also the 
cause for the fraction of mass in substructures to remain consistent; this can happen if the overall 
histories are consistent, even if slightly shifted.

%%%%%%%%%%%%%%%%%%%%%%%%%%%%%%%%%%%%%%%%%%%%%%%%%%%%%%%%%%%%%%%%%%%%%%%%%%%%%%%%%%%%%%%%%%%%%%%%%
%%%%%%%%%%%%%%%%%%%%%%%%%%%%%%%%%%%%%%%%%%%%%%%%%%%%%%%%%%%%%%%%%%%%%%%%%%%%%%%%%%%%%%%%%%%%%%%%%

\section{Tests on varying the extrapolation baseline}
\label{sec:extrapolation}

The method of adjusting the cosmological parameters of a set of haloes and their assembly 
histories can be very useful for exploring a cosmological parameter space using halo properties 
from numerical simulations. In this section we investigate how the agreement between the rescaled 
and reference haloes degrades as the baseline of the extrapolation on the parameter space varies.

We choose to vary only two cosmological parameters for this test, the matter density parameter 
$\Omega_m$, and the amplitude of fluctuations $\sigma_8$.  In this case we take as a starting 
point the parameters of simulation $Blow$,\footnote{The choice of small box and low resolution 
is justified by the precision of the rescaled haloes in the low resolution simulation $Blow$ shown 
in Figure \ref{fig:indiv}, and owes to the large number of simulations involved in this analysis.} 
which are in agreement with the latest constraints on the cosmology \citep{jarosik10}, and explore 
the resulting rescaled haloes for the $Bo_{-2}$, $Bo_{-1}$, $Bo_{+1}$, $Bo_{+2}$ simulations, which 
maintain the same cosmological parameters as simulation $Blow$ but vary $\Omega_m$ between 
$\Omega_m-2\sigma$, and $\Omega_m+2\sigma$ with $\sigma=0.03$, the simulations $Bs_{-2}$, $Bs_{-1}$, 
$Bs_{+1}$, $Bs_{+2}$, which with respect to $Blow$ only change $\sigma_8$ between the values 
$\sigma_8-2\sigma$, and $\sigma_8+2\sigma$ with $\sigma=0.03$, and a simulation in which both, 
$\Omega_m$ and $\sigma_8$ are increased in one standard deviation, $Bo_{+1}s_{+1}$ (details on 
the simulation parameters are shown in Tables 1 and 2). In all cases, the value of $\sigma$ is 
similar to the standard deviation in these parameters from the WMAP7 results. Since we are using 
simulations with small boxes for this test, we only apply the reduced version of the AW10 algorithm 
in this section.

\begin{figure*}
\centering
\includegraphics[scale=0.4]{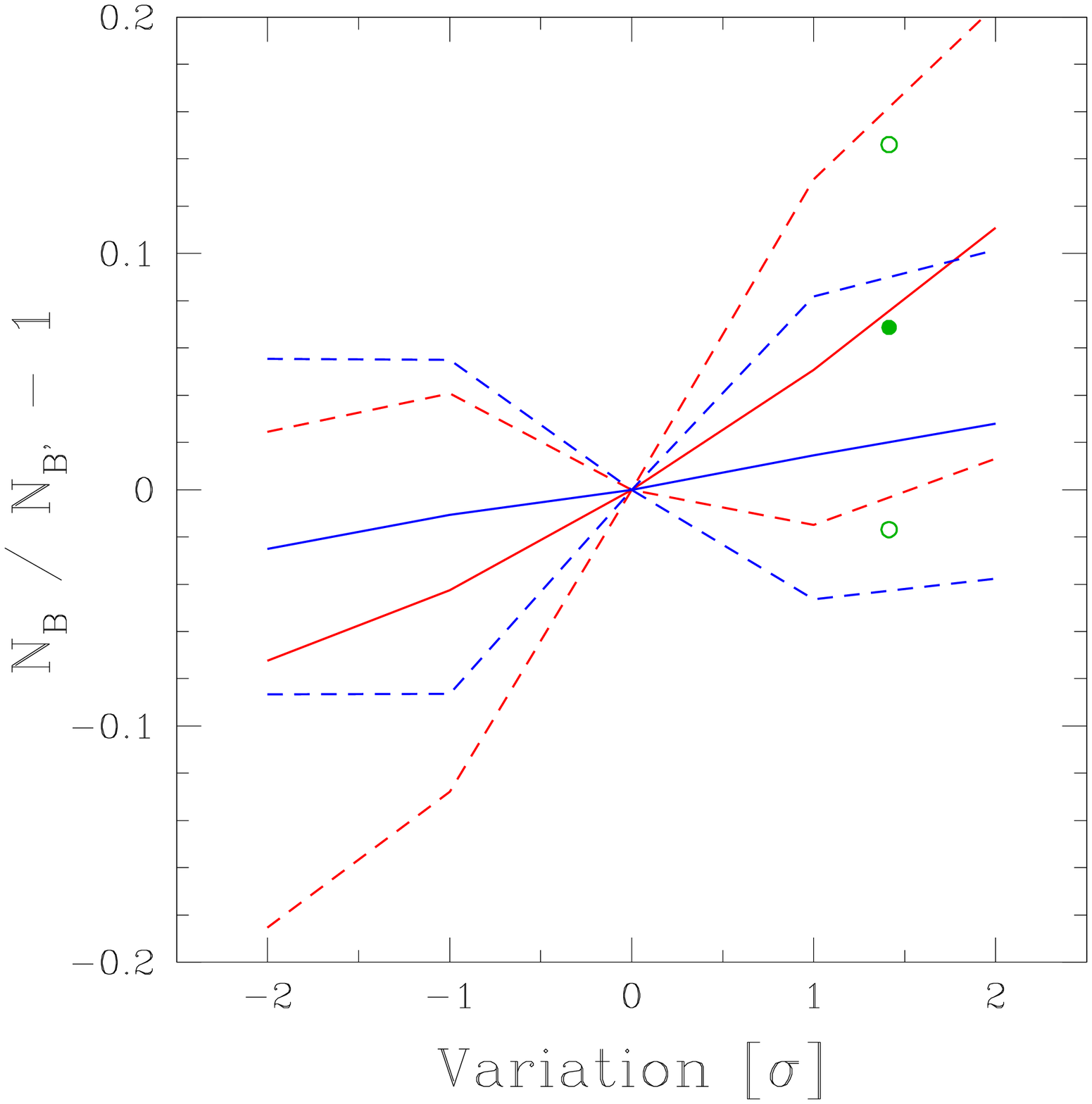}
\includegraphics[scale=0.4]{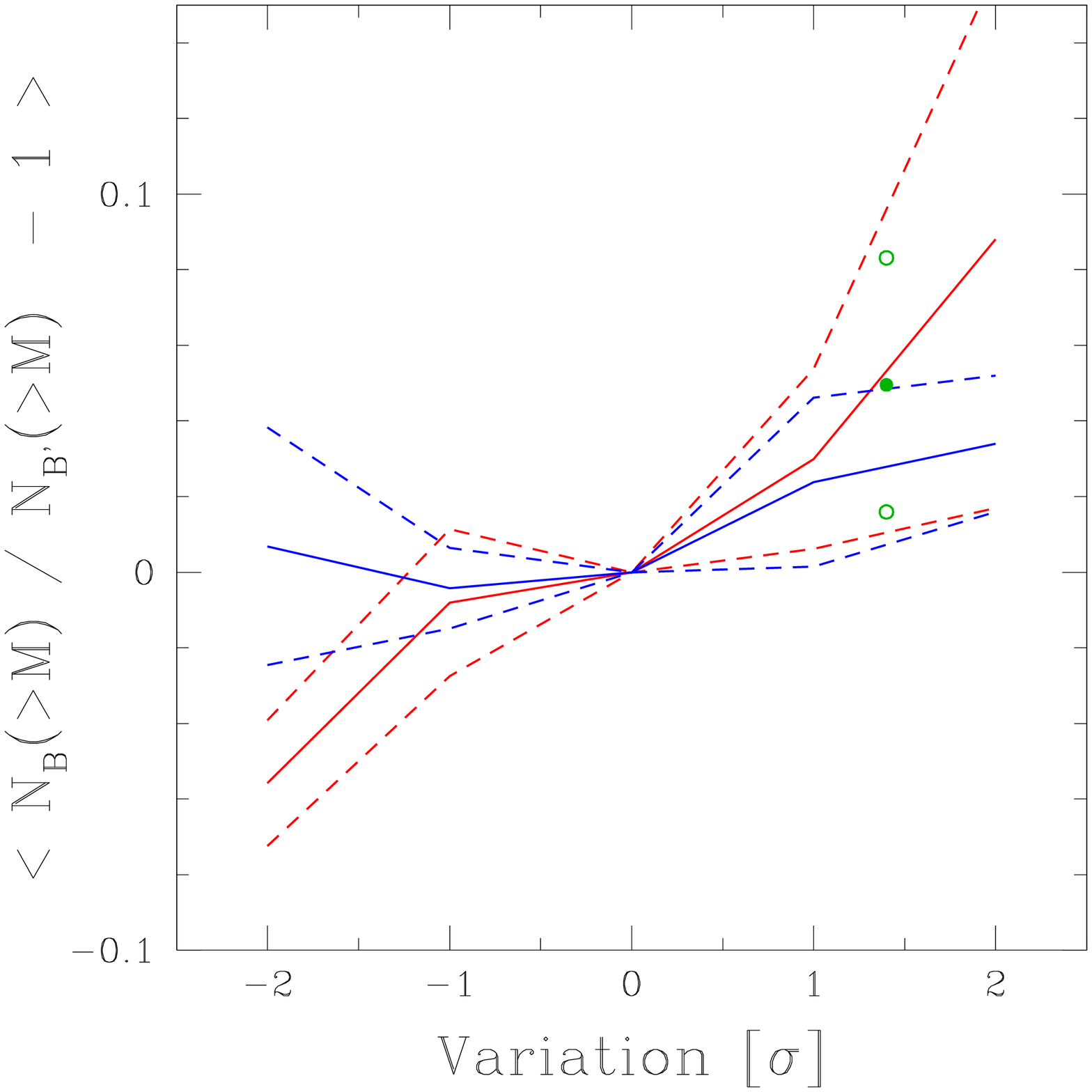}\\
\includegraphics[scale=0.4]{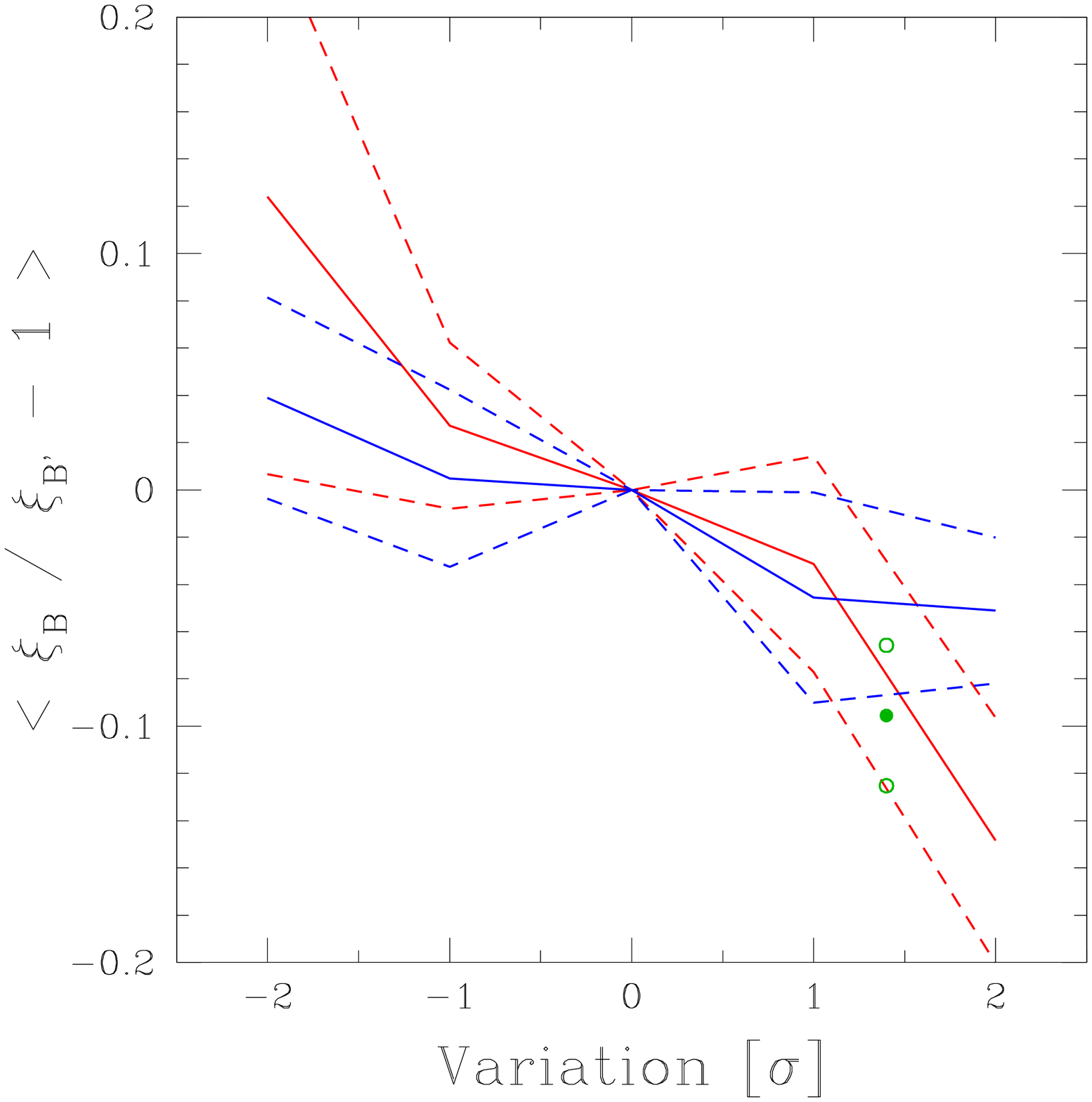}
\includegraphics[scale=0.4]{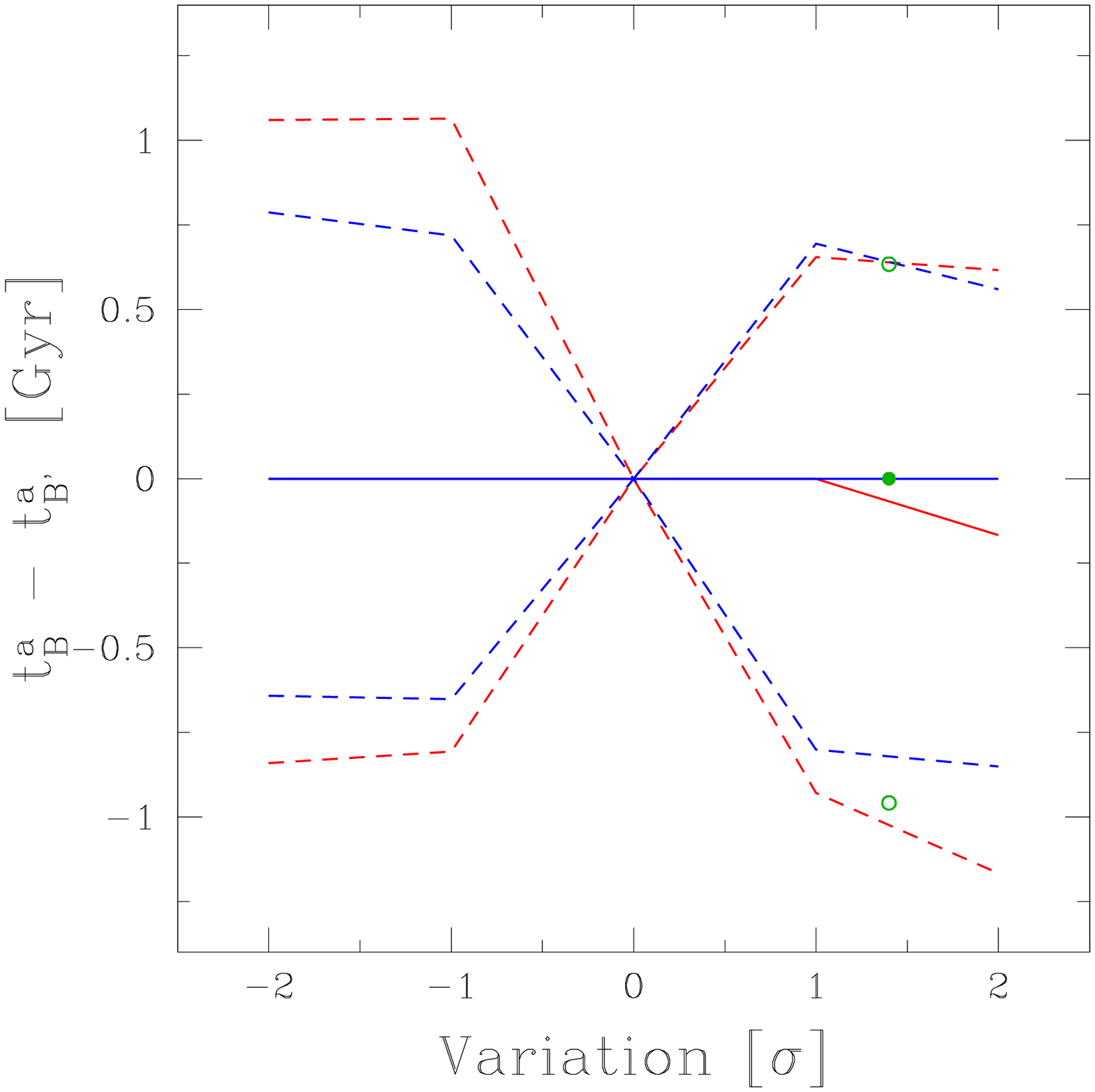}
\caption{
Recovery of halo properties as the baseline for the extrapolation in cosmological parameters
extends away from the base values. In all the panels the blue lines correspond to varying the 
$\sigma_8$ parameter, and the red lines to variations in $\Omega_m$.  Solid lines show the mean 
(top-right and bottom-left) or median values (top-left and bottom-right). The dashed lines show 
the $10$ and $90$ percentiles of the resulting distributions in the upper-left and lower-right,
and the dispersion in the other two panels. The green solid symbol shows the recovery of halo 
properties for the model in which both, $\Omega_m$ and $\sigma_8$ are increased by $1\sigma$, 
shown at $\sqrt{1^2+1^2}\sigma$ on the x-axis; the open symbols show the $10$ and $90$ percentiles, 
or the dispersion. Top-left: variation in the cumulative number of haloes with masses 
$11<\log_{10}(M/h^{-1}M_{\odot})<13$. Top-right: variation in the number of particles per 
dark-matter halo, for haloes with $200$ to $1000$ particles. Bottom-left: variation in the 
correlation functions, averaged on $1<r/$h$^{-1}$Mpc$<10$.  Bottom-right: difference between 
the assembly time-scales $t^a$ of rescaled and reference haloes.}
\label{fig:cosm}
\end{figure*}

Figure \ref{fig:cosm} shows average and median quantities that summarise the accuracy of the rescaled 
haloes. We include the variation in the number of particles per halo averaged over all haloes with 
$200$ to $1000$ particles and the $10$ and $90$ percentiles of the distribution (top-left), the average 
relative variation in the cumulative mass function for masses in the range $11<\log_{10}(M/h^{-1}M_{\odot})<13$ 
and the dispersion of the distribution  (top-right), the average variation in the cross-correlation function 
between haloes of FoF masses $>10^{12}$h$^{-1}M_{\odot}$ against the full halo population, in the range 
of scales $1<r/h^{-1}$Mpc$<10$ and its dispersion (lower-left), and the median difference in the assembly 
ages of rescaled and reference haloes and the $10$ and $90$ percentiles of the distribution (lower-right).
In all cases, the $x-$axis shows the number of WMAP7 standard deviations from the $Blow$ parameters.  
As can be seen, the effect from changing the matter density parameter is in all cases more important than 
varying the amplitude of fluctuations $\sigma_8$. We also show the resulting biases and uncertainties 
in the solid and open green symbols when changing both $\Omega_m$ and $\sigma_8$ by $+1\sigma$, which 
follows the expected tendency and results in larger biases and dispersions than varying only one parameter 
at a time.

These statistics are chosen since they are representative of important variations in the galaxy population that
can be obtained via semi-analytic models (for instance via the abundance of haloes or the assembly time-scales), 
or variations in constraints on cosmological models via clustering and abundance measurements.

The biases in the number of particles per halo and the number of haloes shown in the top panels are 
correlated, and consist on underestimates (overestimates) for lower (higher) values of either 
$\Omega_m$ and $\sigma_8$. This is due to the fact that a bias in the number of particles, which does 
not depend on the halo mass as was seen in the comparison between simulations $A$ and $B$, $Alow$ and 
$Blow$ and $Abig$ and $Bbig$ (cf. Figure \ref{fig:indiv}, although an underestimation of the number 
of particles per halo is present), is reflected in the mass function. The change from simulation $A$ 
to $B$ is approximately equivalent to $-3\sigma$ and $+1\sigma$ in $\sigma_8$ and $\Omega_m$, respectively, 
which would correspond to  $\sqrt{1^2+3^2}\sigma\sim-3\sigma$ in the top-left panel of Figure 
\ref{fig:cosm}, where the number of particles per halo in rescaled catalogues is underestimated, 
showing the consistency of our analysis.

\begin{figure}
\centering
\includegraphics[angle=-90,scale=0.4]{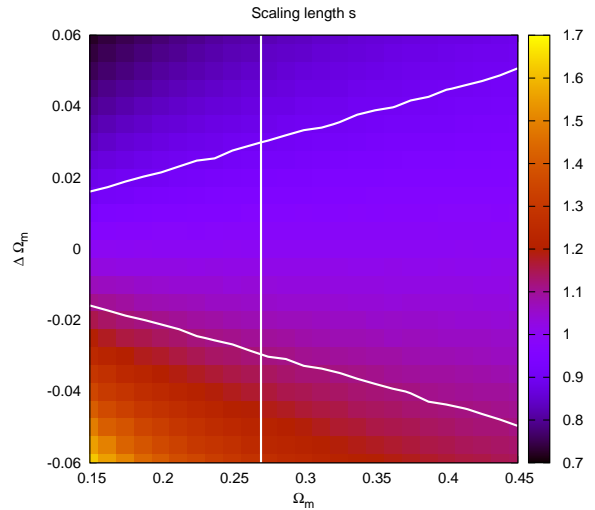}
\caption{Rescaling parameter $s$ (color gradient) as a function of the starting cosmology 
parameter $\Omega_m$ (x-axis) and $\Delta \Omega_m$ needed to reach the desired cosmology 
(y-axis). The remaining cosmological parameters are those of the $B$ simulation, with the exception 
of $\sigma_8$ which is set to $0.82$ for the starting cosmology and $0.73$ for the desired one. 
The white vertical line indicates the position of $\Omega_m=0.27$, and the diagonal lines represent 
the contours crossing $\Omega_m=0.27$ when $\Delta \Omega_m= \pm 0.03$. The scaling length parameter 
$s$ is not sensitive to changes in $\sigma_8$.}
\label{fig:map}
\end{figure}

The lower-left panel of Figure \ref{fig:cosm} shows that the amplitude of the correlation function
shifts from an overestimation when the parameters are lowered, to an underestimation when going to higher
values of either $\Omega_m$ or $\sigma_8$ (with a larger variation in the former). Finally, the asembly 
times shown on the lower-right show almost no biases, but a slight tendency towards underestimating 
this quantity for positive $\sigma$ values on either $\sigma_8$ or $\Omega_m$ (and stronger for the latter).

The increase in number of particles per halo, the space density of haloes, and the decrease in 
the amplitude of the correlation function (and possibly in the assembly time-scales) responds to the 
decreasing values of the rescaling parameter $s$.  This can be seen in Figure \ref{fig:map} where we 
show in a colour gradient that, for a fixed value of $\Omega_m$ (for instance, that of the $Blow$ 
simulation, shown by the vertical line), a higher final value of matter density parameter requires $s<1$. 
This figure also serves the purpose of allowing to rescale the expected biases shown in Figure \ref{fig:cosm}
to a different base cosmology. The white diagonal lines are contours of the equal $s$ values that satisfy
$\Delta \Omega_m=0.03$ at the $Blow$ cosmological parameter set, and show that for lower values of 
base cosmology $\Omega_m$, the biases of Figure \ref{fig:cosm} would only be obtained for smaller 
values of $\Delta \Omega_m$. As this result is independent of the base and desired values of $\sigma_8$, 
this shows that a grid of cosmological simulations would need to more densely cover low $\Omega_m$ values, 
roughly following $\Delta \Omega_m=0.03 \times \left(\Omega_m/0.27 \right)$, to ensure a stable accuracy 
in rescalings done along the $\Omega_m$ parameter, for any value of $\sigma_8$.  As for changes in 
$\sigma_8$, as this will entail changing the output redshift adopted for the rescaling rather than 
introducing a new dimension in the grid of simulations, this will not affect the design of the grid.

These results can be used to set the maximum difference on either $\Omega_m$ or $\sigma_8$ that 
will be allowed when adopting the AW10 method to rescale dark-matter haloes from a given cosmology 
into different ones. This result is also expected to hold for larger simulation boxes as long as the 
full rescaling algorithm is applied.

%%%%%%%%%%%%%%%%%%%%%%%%%%%%%%%%%%%%%%%%%%%%%%%%%%%%%%%%%%%%%%%%%%%%%%%%%%%%%%%%%%%%%%%%%%%%%%%%%
%%%%%%%%%%%%%%%%%%%%%%%%%%%%%%%%%%%%%%%%%%%%%%%%%%%%%%%%%%%%%%%%%%%%%%%%%%%%%%%%%%%%%%%%%%%%%%%%%

\section{Conclusions}
\label{sec:conclusions}

We have tested the use of the full and a reduced form of the Angulo \& White (2010) method to 
change the cosmology of a catalogue of simulated haloes with the aim to test its application to 
a cosmological parameter search using semi-analytic galaxies. Our tests on the effect of applying 
it to the haloes in a numerical simulation instead of to the individual particles, shows a dramatic 
reduction of computational time. The main reason for this particular test is that future planned
numerical cosmological simulations will be so large that the storage of their individual particles
will be impractical, in which case only halo properties will be available. Even in the event that 
the approach presented here is applied to a simulation with available individual particles, the CPU 
time required to rescale DM haloes is orders of magnitude (at least one) smaller than it is required 
to rescale the particles.  If one adds the time consumed in identifying DM haloes and constructing 
merger trees, the speed up of the process can reach an improvement of two or three orders of magnitude 
(R. Angulo, private communication). The reduced form of the rescaling algorithm further reduces the 
computational time, but at the expense of being limited to small simulation boxes. For the cosmological 
parameters of simulations $A$ and $B$ (see Table \ref{tab:scaling}), the simulation box side 
should be $\sim 50$h$^{-1}$Mpc or smaller. In such boxes, the correction for the quasi-linear modes 
does not further improve the accuracy of the rescaling.

We measured the achieved precision of the rescaled haloes extracted from a given set of cosmological 
parameters (such as those adopted in the Millennium simulation) by comparing them to those extracted 
from numerical simulations with the desired final parameters (in a first instance corresponding to a 
Bolshoi simulation cosmology), constructed using the same initial conditions as the simulation from 
which the rescaled haloes are taken. We compared individual properties, such as the number of particles 
per halo, their peculiar velocities (and variation in the direction of movement), concentration parameters, 
the mass of the haloes, and their positions. In all cases the precision of the recovered properties is 
comparable to what is obtained from the full modification of the individual particles in the simulation 
(AW10). The level of precision is better than $100$kpc for the halo positions, and of $5$ percent of 
their peculiar velocities. Both the number of particles per halo, and the virial halo mass are slightly 
underestimated, by $5$ percent. 

It should be noticed, though, that in order to obtain this accuracy for the virial mass, it is necessary 
to perform an additional correction which takes into account the different values of 
$\delta_{vir}(z,\Omega_m)$ in the rescaled and target cosmologies. The halo concentrations 
derived using the \citet{bullock01} recipe shows the lowest biases between rescaled and reference
cosmologies, with no dependence on halo mass.

We tested the effects from lowering the resolution of the simulation by a factor of $8$ in the number 
of particles (for the same total periodic volume), and from increasing by a factor of $\sim 100$ the 
volume of the simulation (for the same number of particles). We expect changes in the two cases due 
to different reasons. In the case of changing the resolution, all the other parameters of the rescaling 
of the haloes were held fixed. In the change of the simulated volume, the range of masses over which 
the rescaling function was minimised was shift one order of magnitude to higher masses (see Table 
\ref{tab:scaling}). Lowering the resolution or increasing the volume do not change the precision in 
the recovered number of particles as a function of the number of particles in the halo. This implies 
that a lower resolution results in a lower precision at a fixed halo mass. The lower resolution does 
not produce any other significant biases. The increase of the volume does not result in a higher 
uncertainty in the recovered positions since the quasi-linear correction of long wavelength 
contributions in the AW10 method properly takes into account the different quasi-linear theory modes 
in the two models.

Furthermore, we also explored the differences arising in the detailed history of growth of haloes,
and find only mild displacements in particular events such as major mergers.  In order to do this
we checked haloes corresponding to the $1$st., 10th, 100th and 1000th most massive objects in the 
simulations, and the resulting agreement is independent of the halo mass, at least down to masses 
$M\simeq8\times 10^{11}$h$^{-1}M_{\odot}$. These results indicate that rescaled haloes and halo 
histories can even be used independently with little effects on the resulting galaxy population.

A direct application of this method consists on changing the cosmological background of semi-analytic
galaxies.  In order to assess the possible systematic biases that such a population of galaxies
would present if it was based on a catalogue of rescaled haloes, we studied the offset in several 
measures of the halo merger histories, on the halo mass funcion and their clustering properties,
on their spin parameter distributions, and in the number of major mergers experienced by the 
haloes.  Neither the mass function nor the correlation amplitudes show important variations 
($<5$ percent) within the range of masses explored, which is also the case for the distribution 
of spin parameters or frequency of major mergers.  The only possible important discrepancy comes 
from a systematic offset in the ages of merger trees (assembly, star formation and last starburst), 
which are all biased toward smaller ages.  Given that these are correlated, the relative shapes 
of the spectral energy distributions of galaxies of different types will remain almost unchanged, 
but will result in a global change of the population toward slightly bluer colours.  

This method can effectively allow to sample the cosmological parameter space using fully 
non-linear simulations. Even though semi-analytic models were designed to understand the 
processes driving galaxy formation and evolution, the possible dependences of galaxy 
properties on the cosmological parameters could also be used to impose constrains on the 
latter. To do this one possible approach is to use Monte-Carlo Markov Chain or similar 
analyses of the cosmological parameter space (see for instance \citealt{harker07} and 
\citealt{bower10}). This would produce chains of parameter sets on which, in principle, 
a new numerical simulation would need to be run, in which the detection of haloes and 
merger trees, and a semi-analytic model, would need to be performed and applied. This 
process becomes much more efficient when using rescaled haloes, which avoid several 
time-consuming processes. The lowest computational cost is achieved by rescaling haloes 
in boxes small enough so as to allow the use of the reduced method.

At this point it is important to note that even if there are small systematic
effects on the scaled catalogues and merger trees, these are negligible in comparison 
to the uncertainties in the semi-analytic modeling.  Furthermore, given the flexibility 
of semi-analytic models, it is in principle possible to diminish the difference between 
rescaled and direct simulations, but in a way that is not related to the scaling formalism; 
therefore, this particular ability does not help to improve the use of rescalings in
a fast search for baryonic and cosmological parameters together.

We showed that in terms of the uncertainties in the matter density parameter $\Omega_m$
and amplitude of fluctuations parametrised by $\sigma_8$, the biases on different
halo properties increase considerably more in the case of adopting more different values
of $\Omega_m$ than of $\sigma_8$.  These constraints can be used to set limits on how far
the extrapolation can be extended in terms of these parameters; this way only a limited 
number of numerical simulations located in strategically selected grid points would need 
to be run in order to cover in a continuous way a wide cosmological parameter space.

%%%%%%%%%%%%%%%%%%%%%%%%%%%%%%%%%%%%%%%%%%%%%%%%%%%%%%%%%%%%%%%%%%%%%%%%%%%%%%%%%%%%%%%%%%%%%%%%%
%%%%%%%%%%%%%%%%%%%%%%%%%%%%%%%%%%%%%%%%%%%%%%%%%%%%%%%%%%%%%%%%%%%%%%%%%%%%%%%%%%%%%%%%%%%%%%%%%

\section*{Acknowledgments}

We have benefited from helpful discussions with the referee Ra\'ul Angulo. ANR acknowledges 
receipt of fellowships from the Consejo Nacional de Investigaciones Cient\'ificas y T\'ecnicas 
(CONICET-Argentina) and FONDAP $15010003$.  NP acknowledges support from Proyecto Fondecyt 
Regular no. 1071006, FONDAP $15010003$, and project Basal PFB0609, and wishes to thank the 
hospitality of the IATE institute at Univeridad Nacional de C\'ordoba during a scientific visit 
to work on this project. MJD acknowledges SeCyT (UNC - Argentina) and CONICET grants. The authors 
kindly acknowledge Volker Springel for providing the {\small SUBFIND} code.  This project made 
use of $\sim 2$ months of CPU time in the Geryon cluster at the Centro de Astro-Ingenier\'\i a UC.

%%%%%%%%%%%%%%%%%%%%%%%%%%%%%%%%%%%%%%%%%%%%%%%%%%%%%%%%%%%%%%%%%%%%%%%%%%%%%%%%%%%%%%%%%%%%%%%%%

\end{document}